\documentclass[prl,aps,twocolumn,preprintnumbers, showpacs, nofootinbib,superscriptaddress,notitlepage]{revtex4-1}
\usepackage{mathrsfs}
\usepackage{amsfonts}
\usepackage{amsmath}
\usepackage{slashed}
\usepackage{array}
\usepackage{verbatim}
\usepackage{epsfig}
\usepackage{graphicx}
\usepackage{color}
\usepackage[dvipsnames]{xcolor}
\usepackage{multirow}
\usepackage{booktabs}
\usepackage{mathrsfs}
\usepackage[normalem]{ulem}

\newcommand{\beq}{\begin{eqnarray}}
\newcommand{\eeq}{\end{eqnarray}}

\begin{document}

\title{Distribution Amplitudes of  $K^*$ and $\phi$  
at Physical Pion Mass from Lattice QCD}

\collaboration{\bf{Lattice Parton Collaboration ($\rm {\bf LPC}$)}}

\author{Jun Hua}
\affiliation{INPAC, Shanghai Key Laboratory for Particle Physics and Cosmology, Key Laboratory for Particle Astrophysics and Cosmology (MOE), School of Physics and Astronomy, Shanghai Jiao Tong University, Shanghai 200240, China}

\author{Min-Huan Chu}
\affiliation{INPAC, Shanghai Key Laboratory for Particle Physics and Cosmology, Key Laboratory for Particle Astrophysics and Cosmology (MOE), School of Physics and Astronomy, Shanghai Jiao Tong University, Shanghai 200240, China}
\affiliation{Shanghai Key Laboratory for Particle Physics and Cosmology, Key Laboratory for Particle  Astrophysics and Cosmology (MOE), Tsung-Dao Lee Institute, Shanghai Jiao Tong University, Shanghai 200240, China}

\author{Peng Sun} 
\email{Corresponding author: 06260@njnu.edu.cn}
\affiliation{Department of Physics and Institute of Theoretical Physics, Nanjing Normal University, Nanjing, Jiangsu, 210023, China}

\author{Wei Wang}
\email{Corresponding author: wei.wang@sjtu.edu.cn}
\affiliation{INPAC, Shanghai Key Laboratory for Particle Physics and Cosmology, Key Laboratory for Particle Astrophysics and Cosmology (MOE), School of Physics and Astronomy, Shanghai Jiao Tong University, Shanghai 200240, China}

\author{Ji Xu}
\affiliation{INPAC, Shanghai Key Laboratory for Particle Physics and Cosmology, Key Laboratory for Particle Astrophysics and Cosmology (MOE), School of Physics and Astronomy, Shanghai Jiao Tong University, Shanghai 200240, China}
\affiliation{School of Physics and Microelectronics, Zhengzhou University, Zhengzhou, Henan 450001, China}

\author{Yi-Bo Yang}
\affiliation{CAS Key Laboratory of Theoretical Physics, Institute of Theoretical Physics, Chinese Academy of Sciences, Beijing 100190, China}
\affiliation{School of Fundamental Physics and Mathematical Sciences, Hangzhou Institute for Advanced Study, UCAS, Hangzhou 310024, China}
\affiliation{International Centre for Theoretical Physics Asia-Pacific, Beijing/Hangzhou, China}

\author{Jian-Hui Zhang}
\affiliation{Center of Advanced Quantum Studies, Department of Physics, Beijing Normal University, Beijing 100875, China}

\author{Qi-An Zhang}
\affiliation{Shanghai Key Laboratory for Particle Physics and Cosmology, Key Laboratory for Particle  Astrophysics and Cosmology (MOE), Tsung-Dao Lee Institute, Shanghai Jiao Tong University, Shanghai 200240, China}

\date{\today}
\begin{abstract} 
We present the first lattice QCD calculation of the distribution amplitudes of longitudinally and transversely polarized vector mesons $K^*$ and $\phi$ using large momentum effective theory. We use the clover fermion action on three ensembles with 2+1+1 flavors of highly improved staggered quarks (HISQ) action, generated by MILC collaboration, at physical pion mass and \{0.06, 0.09, 0.12\} fm lattice spacings, and choose three different hadron momenta $P_z=\{1.29, 1.72, 2.15\}$ GeV. The resulting lattice matrix elements are nonperturbatively renormalized in a hybrid scheme proposed recently. An extrapolation to the continuum and infinite momentum limit is carried out. We find that while the longitudinal distribution amplitudes tend to be close to the asymptotic form, the transverse ones deviate rather significantly from the asymptotic form. Our final results provide crucial {\it ab initio} theory inputs for analyzing pertinent exclusive processes.
\end{abstract}

\maketitle

{\it Introduction.}--Searching for new physics  beyond the standard model (SM) is a primary goal of particle physics nowadays.  A unique possibility of doing so is to investigate flavor-changing neutral current processes which are highly suppressed in the SM. Some prominent examples of such processes include $B\to K^*\ell^+\ell^-$ and $B_s\to \phi\ell^+\ell^-$ decays. Recent experimental analyses by Belle and LHCb collaborations~\cite{Wei:2009zv,Aaij:2015esa,Aaij:2015oid,Aaij:2017vbb,Aaij:2020nrf} have revealed notable tensions between the SM predictions  of such processes and data, and attracted quite considerable theoretical interests (see Refs.~\cite{Capdevila:2017bsm,Buttazzo:2017ixm,Cerri:2018ypt} and many references therein). Various new physics interpretations have been proposed to resolve such tensions, but to firmly establish their existence requires an accurate and reliable theoretical understanding of the dynamics of weak decays.

In the low recoil region (high $q^2$), the $B\to K^*$ and $B_s\to \phi$ form factors can be directly calculated on the lattice (see for instance Refs.~\cite{Horgan:2013hoa,Horgan:2013pva}), {however  these decays at large recoil are also of experimental interests, and for instance  the $P'_5$ anomaly  has attracted many theoretical and experimental attentions~\cite{Descotes-Genon:2012isb,LHCb:2015svh}.} In the latter kinematics region,   decay amplitudes are split into short-distance hard kernels and long-distance universal inputs. The universal inputs that enter include the light-cone distribution amplitudes (LCDAs) of the vector mesons $K^*, \phi$ which, to the leading-twist accuracy, specify the longitudinal momentum distribution amongst the valence quark and antiquark in the meson.  While the hard scattering kernel is perturbatively calculable, the LCDAs can only be extracted from nonperturbative methods or from fits to relevant data.  A reliable knowledge of LCDAs is essential in making predictions on physical observables, and in particular the transition form factors at large recoil can be typically affected by ${\cal O}(10\%)$ by the non-asymptotic terms of LCDAs in light-cone sum rules approach~\cite{Ball:1997rj,Ball:2004rg}.  To date most of the available analyses have made use of estimates based on QCD sum rules~\cite{Ball:2007zt} or Dyson-Schwinger equation~\cite{Gao:2014bca}, but a first-principle description of LCDAs for the vector $(K^*,\phi)$ meson is still missing.

Lattice QCD provides an ideal {\it ab initio} tool to access nonperturbative quantities in strong interaction. Though some lowest moments of the $\rho$ LCDA have been studied in Ref.~\cite{Braun:2016wnx}, a direct calculation of the entire distribution has not been feasible until the proposal of large momentum effective theory (LaMET)~\cite{Ji:2013dva,Ji:2014gla} recently.  This  is realized by simulating on the lattice appropriately chosen equal-time correlations,  and then convert them to the latter through a perturbative matching.   
Since LaMET was proposed, a lot of progress has been achieved in calculating various parton distribution functions~\cite{Ji:2020ect,Cichy:2018mum} (and many references therein) as well as distribution amplitudes for light pseudoscalar mesons~\cite{Zhang:2017bzy,Chen:2017gck,Zhang:2020gaj}. Other variants have also been explored in Refs.~\cite{Ma:2014jla,Ma:2017pxb,Radyushkin:2017cyf}.  

In this Letter, we present the first lattice calculation of LCDAs for vector mesons $K^*,\phi$ in LaMET with the clover fermion action, on three ensembles with 2+1+1 flavors of highly improved staggered quarks (HISQ) action~\cite{Follana:2006rc}, generated by MILC collaboration~\cite{Bazavov:2012xda}, at  physical  pion mass and 0.06, 0.09 and 0.12 fm lattice spacings. To improve the signal-to-noise ratio of simulation, we take the smearing transformation of hyperubi(HYP) fat link \cite{Hasenfratz:2001hp}, the other simulation setup is given in Table I. A hybrid renormalization scheme~\cite{Ji:2020brr} is used to renormalize   bare quantities, after which an extrapolation is taken to the continuum limit as well as to the infinite momentum limit based on data at three hadron momenta, $P_z=\{1.29, 1.72, 2.15\}$ GeV. {It should be noticed that, a momentum boost close to or larger than the inverse lattice spacing may introduce uncontrolled discretization effects, while the dispersion relation is satisfied with the ${\cal O}(a^2)$ corrections as shown in the supplemental material~\cite{supplementary}.} In the calculation we   neglect the strong decays of $K^*,\phi$ due to their narrow decay widths. The finite width corrections should be solved with a proper finite-volume analysis, which is beyond the scope of this work. Our final results indicate that, while the longitudinal LCDAs are close to the asymptotic form, the transverse ones deviate   considerably  from the asymptotic form. 

%%%%%%%%%%%%%
\begin{table}
\centering
\caption{Information on the simulation setup. The light and strange quark mass(both valence and sea quark) of the clover action are tuned such that $m_{\pi}$=140 MeV and $m_{\eta_s}$=670 MeV.}
\label{Tab:setup}
\begin{tabular}{cccccccccc}
\hline
\hline
Ensemble ~~& $a$(fm) & $L^3\times T$   & $c_{\mathrm{SW}}$ & $m_{u/d}$  & $m_{s}$  \\
\hline
a12m130  ~~& 0.12  ~~& 48$\times$~ 64  ~~& 1.05088  ~~&-0.0785         ~~&-0.0191           \\
a09m130  ~~& 0.09  ~~& 64$\times$~ 96  ~~& 1.04239  ~~&-0.0580         ~~&-0.0174           \\ 
a06m130  ~~& 0.06  ~~& 96$\times$192   ~~& 1.03493  ~~&-0.0439         ~~&-0.0191            \\ 
\hline
\end{tabular}
\end{table}

%\begin{table}[tbp]
%\caption{Information on the simulation setup. The light and strange quark mass of the clover action are tuned such that $m_{\pi}$=140 MeV and $m_{\eta_s}$=670 MeV.}
%\begin{tabular}{|c|c|c|c|c|}
%\hline
%Ensemble & $a$(fm) & $L^3\times T$   & $m_{\pi}$(MeV)  & $m_{\eta_s}$(MeV) \\ \hline
%a12m130  & 0.12  & 48$\times$\ \ 64  & 140            & 670               \\ \hline
%a09m130  & 0.09  & 64$\times$\ \ 96  & 140            & 670               \\ \hline
%a06m130  & 0.06  & 96$\times$192 & 140            & 670               \\ \hline
%\end{tabular}
%\end{table} 
%%%%%%%%%%%%%

{\it {LCDAs from LaMET.}}--The leading-twist LCDAs for longitudinally and transversely polarized vector mesons, $ \Phi_{V,L}$ and $\Phi_{V,T}$, are defined as follows~\cite{Ali:1993vd}:
\begin{eqnarray} 
&& \int d\xi^- e^{-ixp^+ \xi^-}\langle 0 | \bar \psi_1(0) n\!\!\!\slash_+ U(0,\xi^-)\psi_2(\xi^-) |V\rangle \nonumber\\
&&= f_{V}n_+\cdot  \epsilon \Phi_{V,L}(x), \\
&& \int d\xi^- e^{-ixp^+ \xi^-}\langle 0 | \bar \psi_1(0) \sigma^{+  {\mu_\perp}} U(0, \xi^-)\psi_2(\xi^-) |V\rangle \nonumber\\
&&= f_{V}^T  [\epsilon^{+}p^{\mu_\perp}-\epsilon^{{\mu_\perp}}p^+]  \Phi_{V,T}(x), 
\end{eqnarray} 
where $U(0,\xi^-) = P{\rm exp}\big[ig_s\int_{\xi_-}^0 ds {n}_+\cdot A(s{n}_+)\big]$ is the gauge-link defined along the minus lightcone direction, $\epsilon$ is the polarization vector of the vector meson, and ${n}_{+}$ is the unit vector along the plus lightcone direction. $f_V$ and $f_V^T$ are the decay constants defined by the local vector and tensor current, respectively. 
Here for $K^*$, $\psi_1$ denotes the strange quark field and $\psi_2$ is the light $u/d$ quark. For the $\phi$ meson, both $\psi_{1,2}$ are strange quark fields. 

According to LaMET, the above LCDAs can be obtained by first calculating the following bare equal-time correlations on the lattice
\begin{align} 
 \langle 0 | \bar \psi_1(0) \gamma^t U(0, z\hat z)\psi_2(z \hat z) |V\rangle&=H_{V,L}(z)\epsilon^t  f_{V}, \\
 \langle 0 | \bar \psi_1(0) \sigma_{\nu\rho} U(0,z\hat z)\psi_2(z \hat z) |V\rangle&=H_{V,T}(z)f_{V}^T  [{\epsilon}_{\nu}p_\rho -{\epsilon}_{\rho}p_\nu ],\nonumber
\end{align} 
where the Lorentz indices in the second line  are chosen as $\{\nu, \rho\}={z,y}$, and the gauge-link $U(0, z\hat z)$ is along the $z$ direction. The quantities $H_{V,\{L,T\}}(z)$ can be renormalized nonperturbatively in an appropriate scheme~\cite{Ji:2020brr,Ishikawa:2016znu,Chen:2016fxx,Green:2017xeu,Stewart:2017tvs,Alexandrou:2017huk,Braun:2018brg}. Here we choose the hybrid scheme~\cite{Ji:2020brr} proposed recently which has the advantage that the renormalization factor does not introduce extra nonperturbative effects at large $z$ which distort the IR property of the bare correlations. This scheme works as follows: At $|z|\le z_S$ where $z_S$ is within the region where the leading-twist approximation is valid, we can choose the RI/MOM scheme~\cite{Stewart:2017tvs} to avoid certain discretization effects (alternative choices include, e.g., the ratio~\cite{Radyushkin:2017cyf} scheme), while for $|z|>z_S$ one applies the gauge-link mass subtraction scheme
\begin{align}\label{eq:hybrid}
 &H_{V}^R(z,a,P_z)=\frac{H_{V}(z,a,P_z)}{Z(z,a)}\theta(z_S-|z|)\nonumber\\
 &\hspace{1em}+H_{V}(z,a, P_z)e^{-\delta m(\tilde{\mu})z}Z_{\rm hybrid}(z_S, a)\theta(|z|-z_S),
\end{align}
where the superscript $R$ denotes the renormalized quantity, $\tilde \mu$ denotes the intrinsic scale dependence of the gauge-link including both UV and IR. We have chosen $Z(z,a)$ as the RI/MOM renormalization factor computed from
\begin{align} \nonumber 
Z(z,a) &= \frac{1}{12}{\rm Tr} \Big[ \big<S(p)\big>^{-1} \times \big<  S(p|z)\big> \gamma_z \gamma_5   \\ 
  &\times    \prod_n U_z(n\hat z) S(p|0) \big< S(p) \big>^{-1} \gamma_z \gamma_5 \Big]_{p^2=-\mu_R^2, \atop p_z  =0}. 
\end{align} 
The $Z_{\rm hybrid}$ denotes the endpoint renormalization constant which can be determined by imposing a continuity condition at $z=z_{\rm S}$,
\begin{align}
Z_{\rm hybrid}(z_{\rm S},a) = e^{\delta m(\tilde \mu) z_{\rm S}}/{Z(z_{\rm S},a)  }\,.
\end{align}
The mass counterterm $\delta m(\tilde\mu)$ can be extracted from the RI/MOM renormalization factor~\cite{Ji:2020brr}. The $z_S$ are chosen as $0.24fm$ and $0.36fm$ within perturbative region,  and their difference is treated as a systematic error. 
%\Blue{It is expected that $Z_{\rm hybrid}$ has some dependence on scale $\tilde u$.} 

%It is expected that $Z_{\rm hybrid}$ has some dependence on $z_S$, as made manifest in the above equation, and one needs to examine the stability of the final result with respect to the choice of $z_S$.
%beyond $z_S$ we use the Wilson line mass subtraction scheme~\cite{Ishikawa:2016znu,Chen:2016fxx,Green:2017xeu}. A more detailed discussion is given in the supplemental material. 

By Fourier transforming $H^R_{V,\{L,T\}}$ to momentum space, we then obtain the quasi-DAs
\begin{align} 
 \tilde  \Phi_{V,\{L,T\}}(y, P_z)& = \int dz e^{-iyP_z z} H^R_{V,\{L,T\}}(z, P_z),  \label{eq:quasi_l}
%H_{V,L}(z)\epsilon^t  f_{V}&= \langle 0 | \bar \psi_1(0) \gamma^t U(0, z\hat z)\psi_2(z \hat z) |V\rangle, \nonumber\\
%\tilde \Phi_{V,T}(y)& = \int dz e^{-iyP_z z} H_{V,T}(z), \label{eq:quasi_t}\\
% H_{V,T}(z)f_{V}^T  [{\epsilon}_{\nu}p_\rho -{\epsilon}_{\rho}p_\nu ]&=\langle 0 | \bar \psi_1(0) \sigma_{\nu\rho} U(0,z\hat z)\psi_2(z \hat z) |V\rangle,\nonumber
\end{align} 
where   the continuum limit has been taken. It can be factorized into the LCDAs through the factorization theorem~\cite{Ji:2015qla}:
\begin{align}
&\tilde \Phi_{V,\{L,T\}}(y, P_z, \mu_R) \nonumber\\
&=\int_{0}^1 dx\, C_{V,\{L,T\}}(x,y, P_z, \mu_R, \mu)\Phi_{V,\{L,T\}}(x,\mu), 
\end{align}  
where the matching kernel $C_{V,\{L,T\}}$ was derived first in the transverse momentum cutoff scheme in Ref.~\cite{Xu:2018mpf} and then in the RI/MOM scheme in Ref.~\cite{Liu:2018tox}. The $\mu$ and $\mu_R$ reflect the generic renormalization scale dependence of LCDAs and quasi-DAs. {The matching formula and more details of the hybrid scheme can be found in the supplemental material~\cite{supplementary}.}

{\it {Numerical setup.}}--On the lattice, one directly calculates the  two-point correlation function defined as: 
\begin{align}
 C_{2}^{m}(z,\vec P, t)  &= \int d^3y e^{-i\vec P\cdot \vec y} \nonumber\\
 &\times \langle 0 |\bar \psi_1(\vec y, t) \Gamma_1 U(\vec y, \vec y+ z\hat z)\psi_2(\vec y+ z\hat z, t) \nonumber\\
 &\times \bar  \psi_2(0,0) \Gamma_2 \psi_1(0,0)|0\rangle, 
 \label{eq:2pt}
\end{align} 
where 
the longitudinal polarization case ($m=L$) has $\Gamma_1=\gamma_t$, and $\Gamma_2=\gamma_z$, and  the transverse polarization case ($m=T$) has $\Gamma_1=\sigma_{zy}$, and $\Gamma_2=\gamma_x/\gamma_y$.  
%A ratio of the two-point correlation gives the quasi distribution amplitude:
%\begin{eqnarray}
%\frac{ C_{2}(z,t)  }{ C_{2}(z=0,t)  } = \frac{ H_{V}(z)(1+\Delta C_1 e^{-\Delta Et}) }{(1+\Delta C_2 e^{-\Delta Et}) },
%\end{eqnarray}
%where  $\Phi_{V}(z)$ is the quasi DA in the coordinate space.  $\Delta E$ is the energy difference between the ground state and the first contributing  excited state.   The contributions from   excited state are characterized by the   dimensionless coefficients, $\Delta C_1$ and $\Delta C_2$.   
%For  the practical use,  an additional renormalization is required for the non-local operator, arising from the self-energy of gauge-link and the vertex corrections. 
Then the quasi-DAs can be extracted from the following parameterization:
\begin{eqnarray}
\frac{ C^{m}_{2}(z,\vec P, t)  }{ C^{m}_{2}(z=0,\vec P, t)  } = \frac{ H_{V,m}^{b}(z)(1+ c_{m}(z) e^{-\Delta Et}) }{(1+ c_{m}(0) e^{-\Delta Et}) }, \label{eq:c2_ratio}
\end{eqnarray}
where $c_{m}(z)$ and $\Delta E$ are free parameters accounting for the excited state contaminations, and $H_{V,m}^{b}(z)$ is the bare matrix elements for quasi-DA. When $t$ is large enough, the excited state contaminations parameterized by $c_{m}(z)$ and $\Delta E$ are suppressed exponentially, and the ratio defined in Eq.~\eqref{eq:c2_ratio} approaches the ground state matrix element $H_{V,m}^{b}(z)$. Based on the comparison between the joint two-state fit and constant fit shown in the supplemental material~\cite{supplementary}, we choose to use the constant fit in the range of $t\geq0.54~$fm to provide a conservative error estimate in the following calculation.

The numerical simulation is based on three ensembles with 2+1+1 flavors of HISQ~\cite{Follana:2006rc}  at physical pion mass with 0.06, 0.09 and 0.12 fm lattice spacings. The momentum smeared grid source~\cite{Yang:2015zja} with the source positions $(x_0+j_xL/2,y_0+j_yL/2,z_0+j_zL/2)$ are used in the calculation, where $(x_0,y_0,z_0)$ is a random position and $j_{x,y,z}=0/1$. It allows us to obtain the even momenta in unit of $2\pi/L$ with $\sim8$ times of the statistics. We also repeat the calculation at 8, 6, 4 time slices and fold the data in the normal and reversed time directions, which is equivalent to having $570\times 8 \times 8 \times 2$, $730\times 8 \times 6 \times2$ and $970 \times 8 \times 4 \times2$ measurements at three ensembles at $a=$0.06, 0.09 and 0.12 fm, respectively. 
We have further reversed the $\hat z$ direction in Eq. (\ref{eq:2pt}) to double the statistics.   {Based on the numerical results, we confirmed that the dispersion relation can be satisfied for all the cases up to the ${\cal O}(a^2p^4)$ correction, and the continuum extrapolation in the coordinate space or momentum space provides consistent results~\cite{supplementary}.}

\begin{figure}[!th]
\begin{center}
\includegraphics[width=0.45\textwidth]{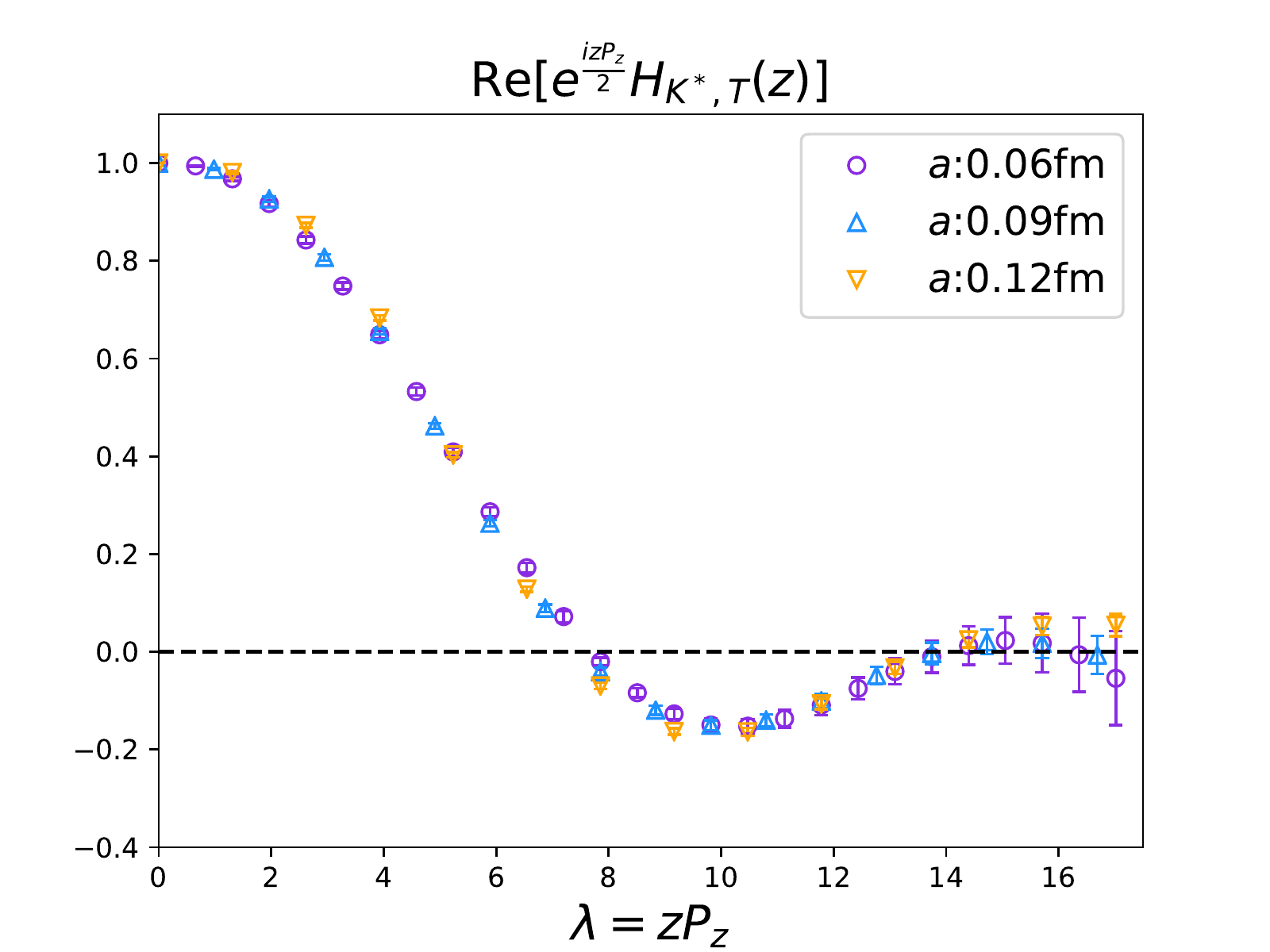}
\includegraphics[width=0.45\textwidth]{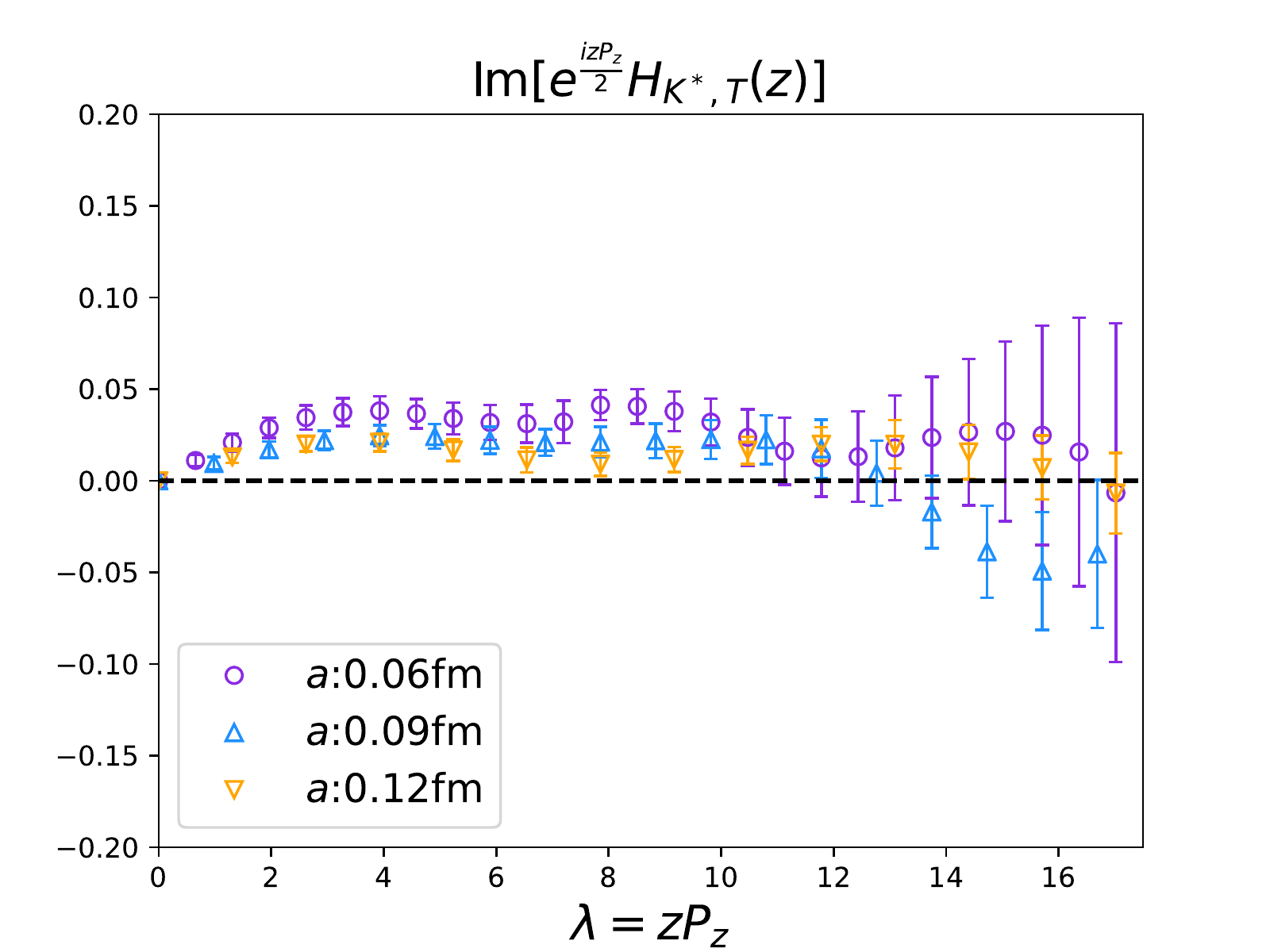}
\caption{The two-point correlation function for the transversely-polarized $K^*$ in coordinate space. We make a phase rotation by multiplying a factor $e^{izP_z/2}$ with $P_z=2.15$GeV. 
}\label{fig:p_rotated_matrix}
\end{center}
\end{figure}

{\it Results.}--After renormalization in the hybrid scheme, we perform a phase rotation $e^{izP_z/2}$ to the renormalized correlation, so that the imaginary part directly reflects the  flavor  asymmetry between the strange and up/down quarks. Taking the transversely-polarized 
$K^*$ as an example, we show in Fig.~\ref{fig:p_rotated_matrix} the real (upper panel) and imaginary part (lower panel) of the renormalized quasi-DA matrix elements $e^{izP_z/2}H_{K^*,T}(z)$ with the momentum $P_z=2\pi /L \times 10 =2.15$ GeV. As shown in the upper panel, the matrix elements at different lattice spacings are consistent with each other, indicating that  linear divergences arising from the gauge-link have been canceled up to the current numerical uncertainty. In the lower panel, we find a positive imaginary part at all the lattice spacings, which corresponds to a non-zero asymmetry with the peak at $x<1/2$. This is consistent with expectations that lighter quarks carry less momentum of the parent meson.

As one can see from Fig.~\ref{fig:p_rotated_matrix}, the uncertainty of lattice data grows rapidly with the spatial separation of the nonlocal operator. %On the other side, in the large $z$ region, some nonperturbative   effects are  uncontrollable in the extraction of LCDAs, which makes a direct use of Lattice data problematic. 
Thus, to have a reasonable control of uncertainties in the final result we need to truncate the correlation at certain point. The missing long-range information can be supplemented by a physics-based extrapolation proposed in Ref.~\cite{Ji:2020brr}, which removes unphysical oscillations in a naive truncated Fourier transform with the price of altering the endpoint distribution (at $x\sim 0$ or $1$), which cannot be reliably predicted by LaMET anyway due to increasingly important higher-twist contributions. %that this problem can be circumvented  by considering the analytical constraints. 
Following Ref.~\cite{Ji:2020brr},  we adopt the following extrapolation form:
\begin{align}
	  H_{V,\{L,T\}}(z, P_z) &= \Big[\frac{c_1}{(-i\lambda)^a} + e^{i\lambda}\frac{c_2}{(i \lambda)^b}\Big]e^{-\lambda/\lambda_0},
	  \label{eq:extra}
\end{align} 
where the exponential term accounts for the finite correlation length for a hadron at finite momentum, and the two algebraic terms account for a power law behavior of the momentum distribution at $x$ close to 0 and 1, respectively. $\lambda = z P_z$, and the parameters $c_{1,2}, a,b, \lambda_0$ are determined by a fitting to the lattice data in the region where it exhibits an exponential decay behavior. To account for systematics from such an extrapolation, we have done two different extrapolations, one including the exponential term and the other not, and taken their difference as an estimate of systematics. {This can be attributed  as a source of the uncertainty from the inverse problem, and a more systematic strategy to handle the inverse problem of the Fourier transform is available  in Ref.~\cite{Karpie:2018zaz}.} The detailed comparison of two extrapolations can be found in the supplemental material~\cite{supplementary}.  

\begin{figure}[!th]
\begin{center}
\includegraphics[width=0.45\textwidth]{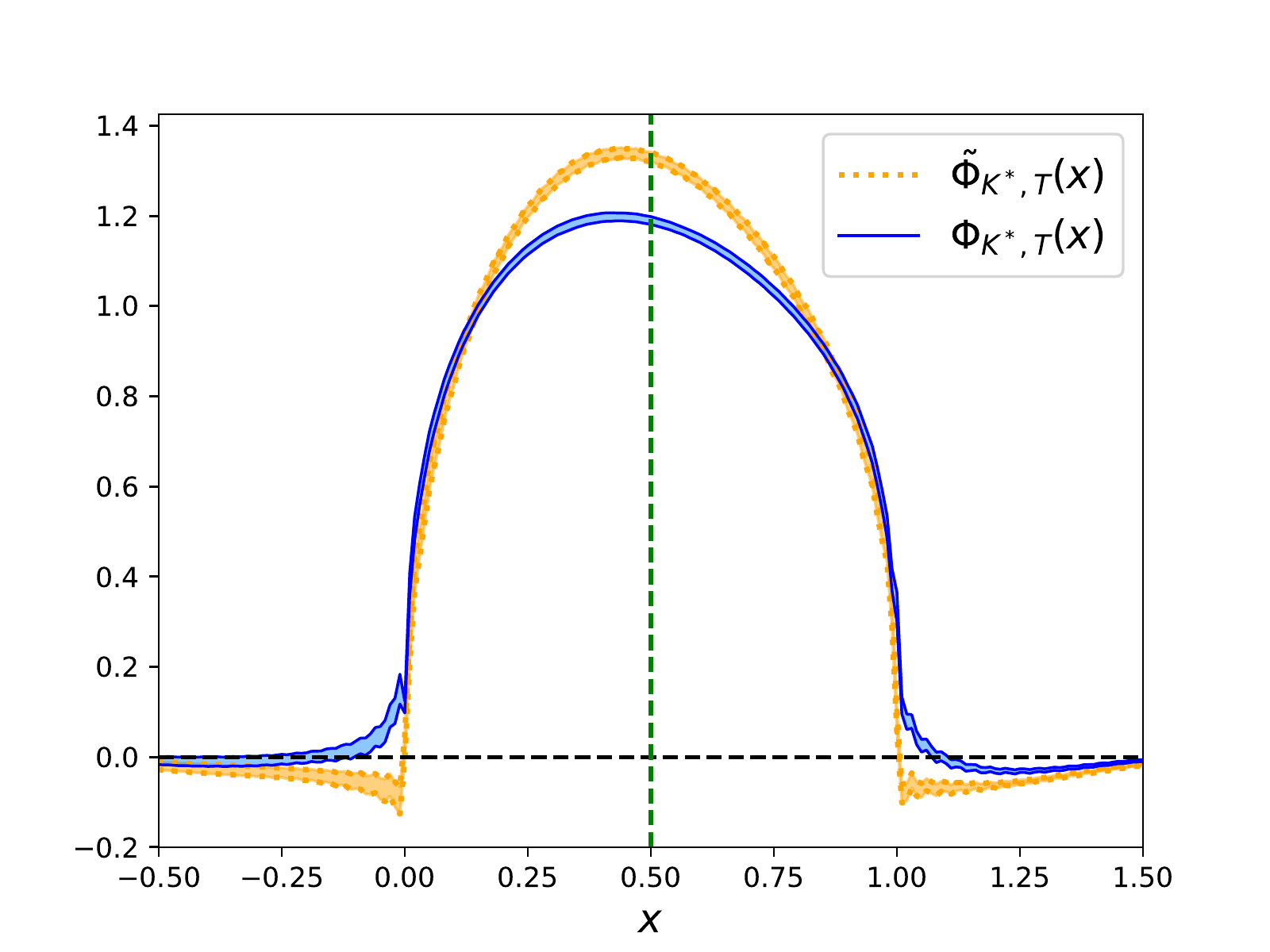}
\caption{Quasi-DA and LCDA extracted from it for the transversely-polarized $K^*$ using data at $a=0.09~$fm, $P_z=2.15$ GeV.}\label{fig:matching_comparison}
\end{center}
\end{figure}
%%%%%%%%%%%%%%%%%%%

After renormalization and extrapolation, we can Fourier transform to momentum space and apply the corresponding matching. In Fig.~\ref{fig:matching_comparison}, we show as an example the comparison of the quasi-DA and extracted LCDA for the transversely polarized $K^*$. The results correspond to the case with $P_z=2.15$ GeV, and  $a=0.09~$fm. One notices that there is a non-vanishing tail for quasi-DA (yellow curve) in the unphysical region ($x>1$ or $x<0$), but it becomes much better for the LCDA (blue curve) after the perturbative matching is applied.

%%%%%%%%%%%%%%%%%%%
%%%%%%%%%%%%%%%%%%%

%%%%%%%%%%%%%%%%%%%

\begin{figure}[!th]
\begin{center}
\includegraphics[width=0.45\textwidth]{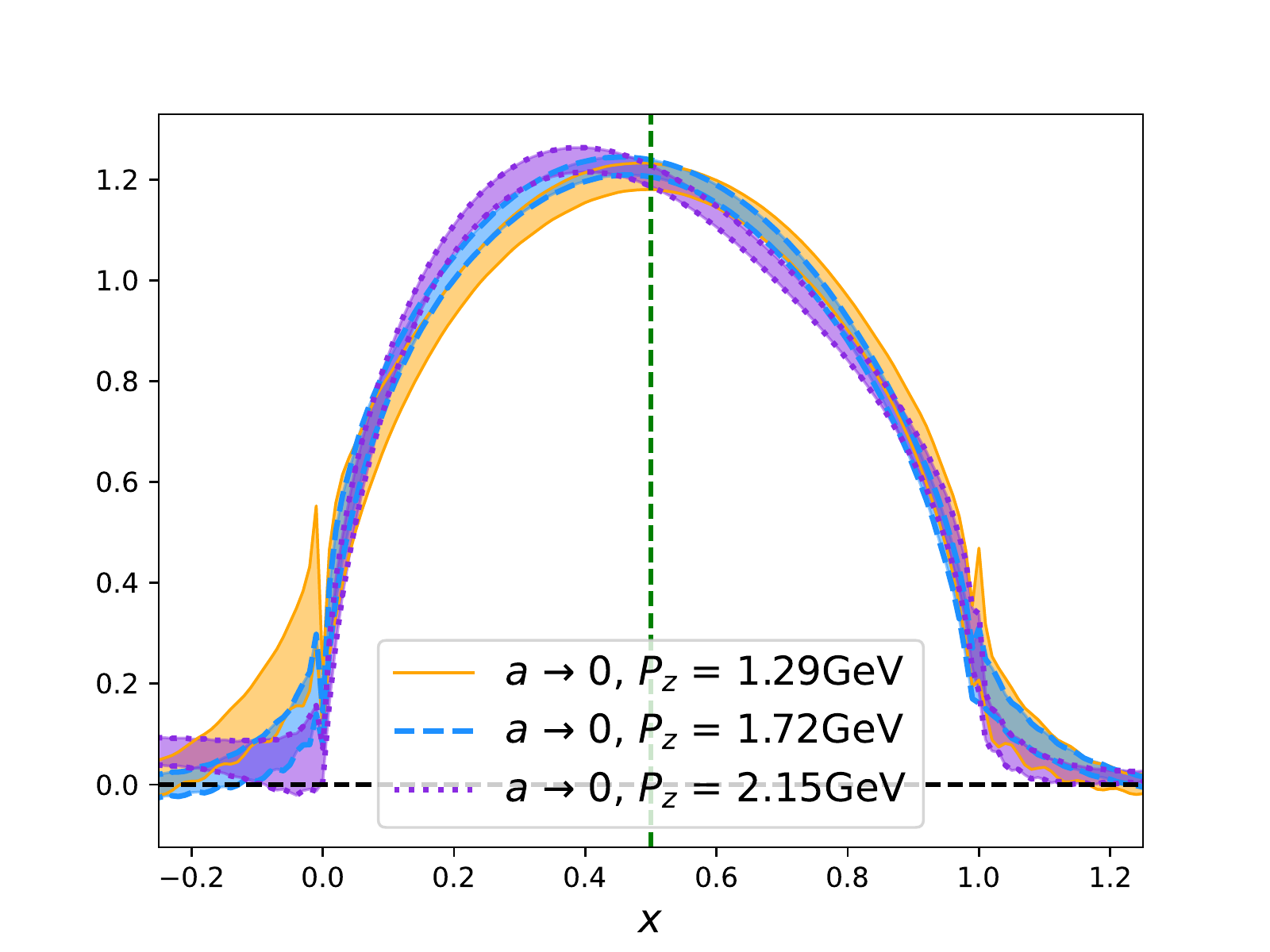}
\caption{The continuum limit of the LCDA for the transversely-polarized $K^*$, extrapolated from three different lattice spacings.}\label{fig:LCDA_continuous_limit}
\end{center}
\end{figure}

We have performed a simple extrapolation to the continuum limit using the results at three different lattice spacings and the following formula
\begin{eqnarray}
	\psi(a) = \psi(a\to0)  + c_1a  + \mathcal{O}(a^2),
\end{eqnarray}
with $\mathcal O(a)$ correction being due to the mixed action effect from the clover valence fermion on HISQ sea. As an example, we show the extrapolated results for the transversely polarized $K^*$ in Fig.~\ref{fig:LCDA_continuous_limit} for three different momenta, $P_z=\{1.29, 1.72, 2.15\}$ GeV. From this figure, one can see that the asymmetry slightly increases with $P_z$. Defining the asymmetry as $c_{asy} = \int^{1/2}_0 dx \phi(x) / \int^1_{1/2} dx \phi(x)$, we find $c_{asy}$ is 1.090(15), 1.176(07), 1.227(08) for the three momenta. Since the strange quark is heavier than the up/down quark, a slight preference of $x<1/2$ to $x>1/2$ is expectable. 
It suggests that a large $P_z$ extrapolation is essential to suppress the power corrections and reproduce this correct preference behavior. Such a behavior is also observed in the study of Kaon LCDAs in Ref.~\cite{Zhang:2020gaj}. 

%%%%%%%
\begin{figure}[!th]
\begin{center}
\includegraphics[width=0.45\textwidth]{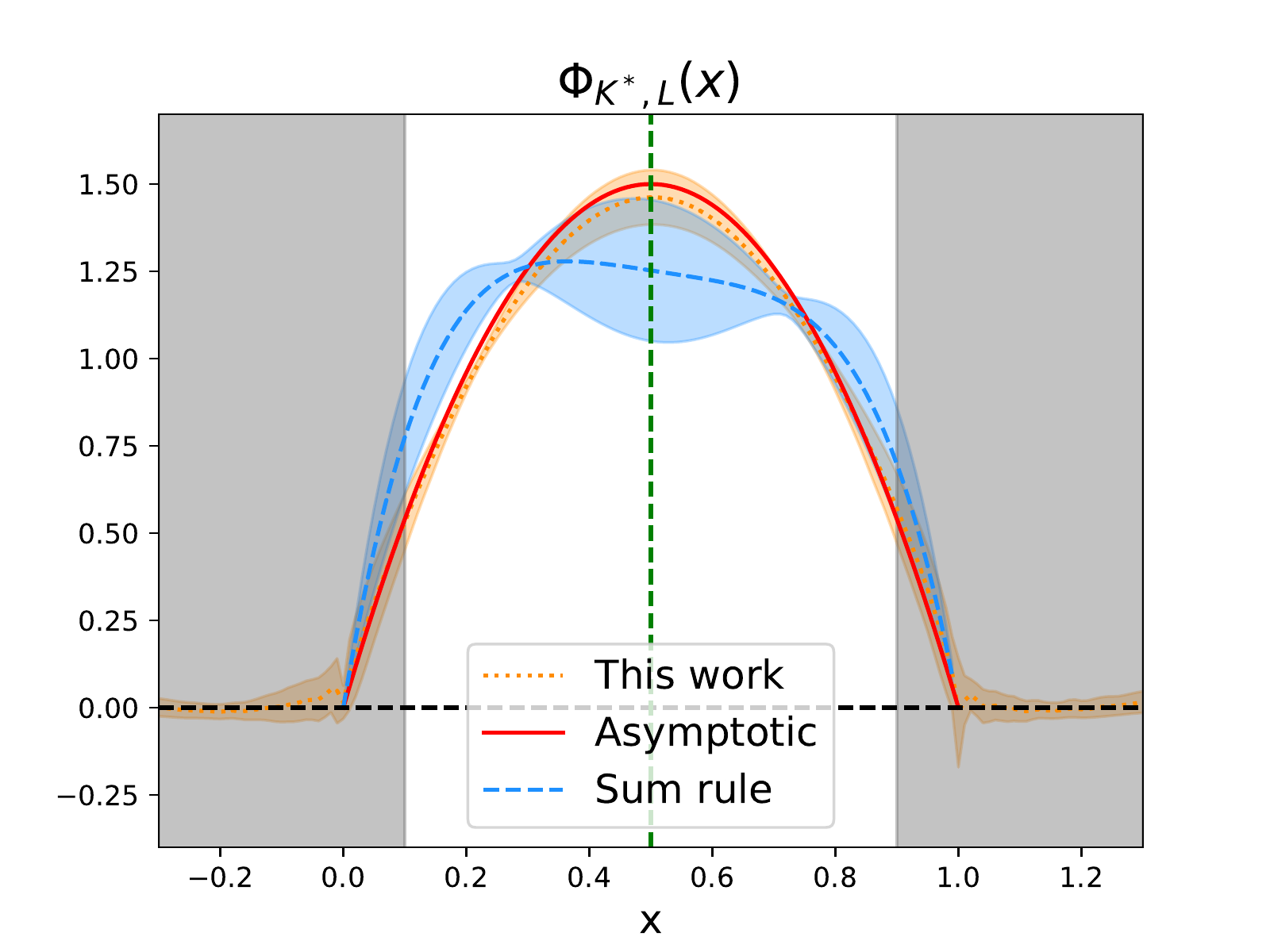}
\includegraphics[width=0.45\textwidth]{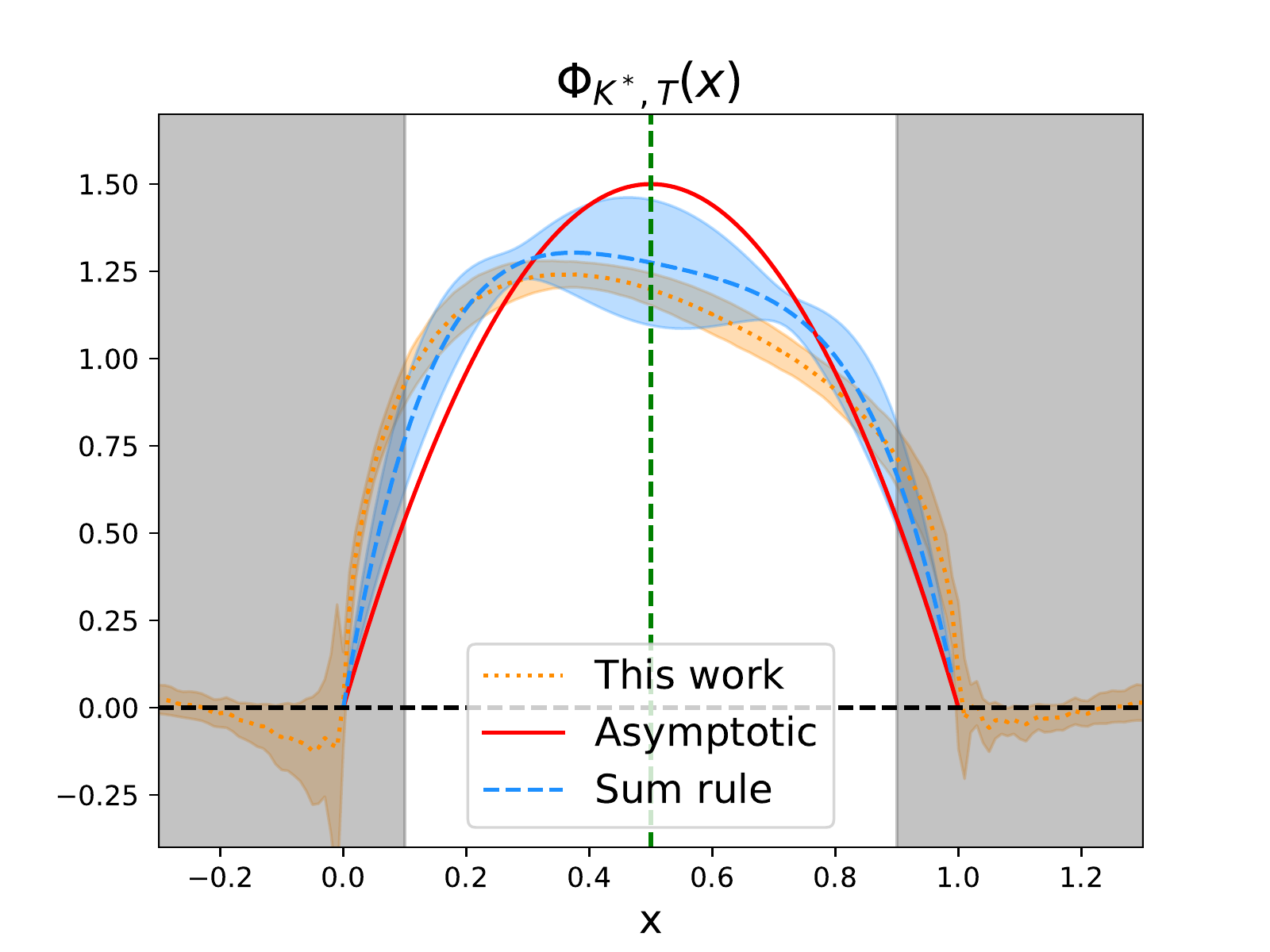}
\caption{LCDAs for the longitudinally-polarized $K^*$(upper panel) and transversely-polarized $K^*$(lower panel). The results are extrapolated to the continuous  limit ($a\to0$) and the infinite momentum limit ($P_z\to\infty$). Regions with $x<0.1, x>0.9$ are shaded, as systematic errors in these regions are difficult to estimate.}\label{fig:LCDA_kstar}
\end{center}
\end{figure}

\begin{figure}[!th]
\begin{center}
\includegraphics[width=0.45\textwidth]{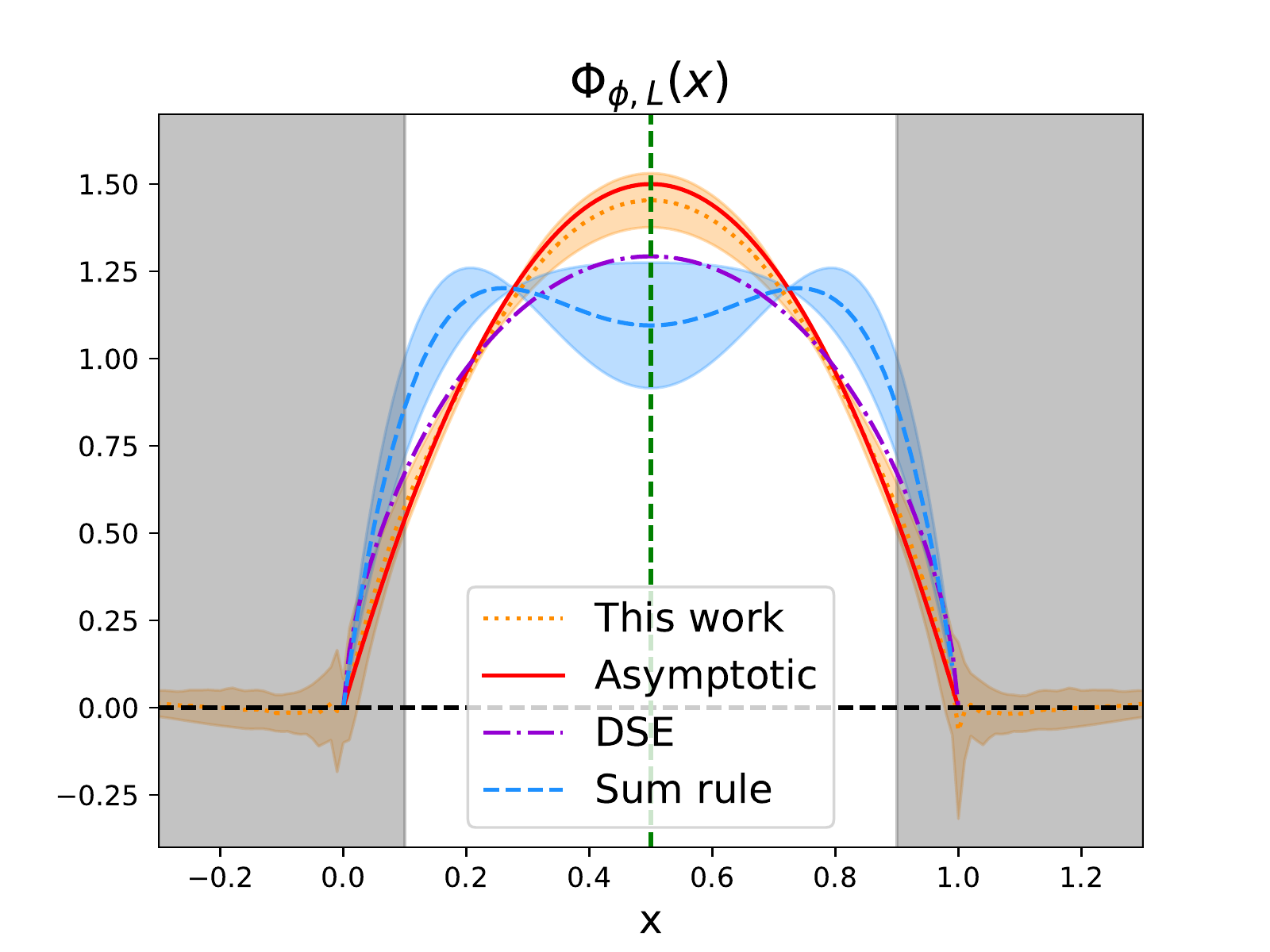}
\includegraphics[width=0.45\textwidth]{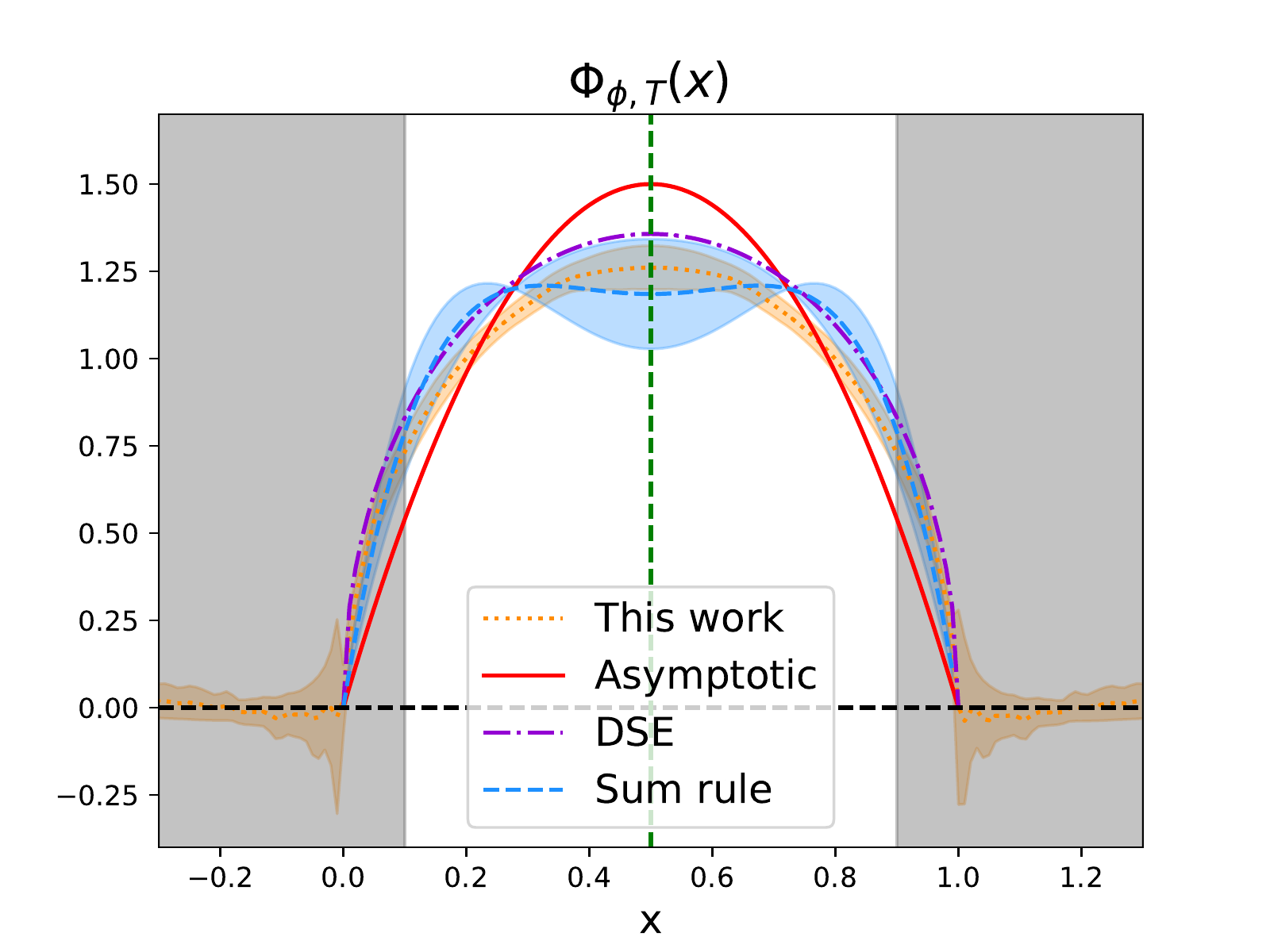}
\caption{Similar to Fig.~\ref{fig:LCDA_kstar}, but for the $\phi$ vector meson.  }\label{fig:LCDA_phi}
\end{center}
\end{figure}

After matching from Quasi-DA to LCDAs with $\mu = 2$ GeV, our final results for LCDAs of the $K^*$ and $\phi$ are given in Fig.~\ref{fig:LCDA_kstar} and \ref{fig:LCDA_phi}, respectively,  where the upper and lower panels correspond to the longitudinal and transverse polarization cases. In these figures, we have made a $P_z\to \infty$ extrapolation using the following simple form:
\begin{eqnarray}
	\psi(P_z) = \psi(P_z \to \infty)  + \frac{c_2}{P^2_z} +   \mathcal{O}\Big(\frac{1}{P^4_z}\Big).
\end{eqnarray}
We have chosen two renormalization scale (1.82GeV and 3.04GeV) and treated their difference an estimate of systematic error from matching. It is worth emphasizing that the endpoint regions are difficult to access in LaMET. The endpoint  can be roughly estimated from the largest attainable $\lambda$ (conjugate variable of $x$ in the Fourier transform) as $1/\lambda_{\rm max}$.   In the present calculation, we have $\lambda_{\rm max}\approx 14$ (specifically $z^{P_z=1.29GeV}_{max}\approx2.1fm$, $z^{P_z=1.72GeV}_{max}\approx1.6fm$, $z^{P_z=2.15GeV}_{max}\approx1.3fm$), thus we take a conservative estimate of the predictable region as $[0.1, 0.9]$. Beyond this region,  we   plot a shaded area with    systematic errors difficult to estimate. As a comparison, we also show in Fig.~\ref{fig:LCDA_kstar} the asymptotic form $6x(1-x)$, the model results  from earlier QCD sum rule calculations~\cite{Ball:2007zt} and the Dyson-Schwinger equations(DSE) results~\cite{Gao:2014bca} in Fig.~\ref{fig:LCDA_phi}. Our results indicate that while the longitudinal LCDAs tend to be close to the asymptotic form, the transverse LCDAs have relatively large deviations from the asymptotic form.  These behaviors might have important implications on the study of semileptonic $B\to K^*\ell^+\ell^-$ decay towards the search for new physics, and can be explored in the future. 

{\it Summary.}--We have presented the first lattice QCD calculation of LCDAs of longitudinally and transversely polarized vector mesons $K^*,\phi$ using LaMET. We did not consider the $\rho$ meson due to its large width which will introduce sizable systematic errors. The continuum and infinite momentum limits are taken based on calculations at physical light and strange quark mass with three lattice spacings and momenta. %After making an extrapolation to the infinite momentum limit, 
Our final results are then compared to the asymptotic form and QCD sum rule results. While the longitudinal LCDAs tend to be close to the asymptotic form, the transverse ones have relatively large deviations from the asymptotic form. Our final results provide crucial {\it ab initio} theory inputs for analyzing pertinent exclusive processes.

{\it Acknowledgement.}--We thank Xiangdong Ji, Liuming Liu, Maximilian Schlemmer, and Andreas Sch\"afer for valuable discussions. 
We thank the MILC collaboration for providing us their HISQ gauge configurations. 
The LQCD calculations were performed using the Chroma software suite~\cite{Edwards:2004sx} and QUDA~\cite{Clark:2009wm,Babich:2011np,Clark:2016rdz} through HIP programming model~\cite{Bi:2020wpt}.
The numerical calculation is supported by Strategic Priority Research Program of Chinese Academy of Sciences, Grant No. XDC01040100. 
The setup for numerical simulations was conducted on the $\pi$ 2.0 cluster supported by the Center for High Performance Computing at Shanghai Jiao Tong University,  HPC Cluster of ITP-CAS, and Jiangsu Key Lab for NSLSCS. 
JH, WW, and JX are  supported in part by Natural Science Foundation of China under grant No. 11735010, 11911530088, U2032102, 11653003,  by Natural Science Foundation of Shanghai
under grant No. 15DZ2272100.
 JH is supported by NSFC under grant 11947215. PS is supported by Natural Science Foundation of China under grant No. 11975127 as well as Jiangsu Specially Appointed Professor Program. YBY is also supported by the Strategic Priority Research Program of Chinese Academy of Sciences, Grant No. XDB34030303. JHZ is supported in part by National Natural Science Foundation of China under Grant No. 11975051, and by the Fundamental Research Funds for the Central Universities.

\end{document}

\title{Supplementary Materials: Distribution Amplitudes of  $K^*$ and $\phi$  
at Physical Pion Mass from Lattice QCD}

\collaboration{\bf{Lattice Parton Collaboration ($\rm {\bf LPC}$)}}

\author{Jun Hua}
\affiliation{INPAC, Shanghai Key Laboratory for Particle Physics and Cosmology, Key Laboratory for Particle Astrophysics and Cosmology (MOE), School of Physics and Astronomy, Shanghai Jiao Tong University, Shanghai 200240, China}

\author{Min-Huan Chu}
\affiliation{INPAC, Shanghai Key Laboratory for Particle Physics and Cosmology, Key Laboratory for Particle Astrophysics and Cosmology (MOE), School of Physics and Astronomy, Shanghai Jiao Tong University, Shanghai 200240, China}
\affiliation{Shanghai Key Laboratory for Particle Physics and Cosmology, Key Laboratory for Particle  Astrophysics and Cosmology (MOE), Tsung-Dao Lee Institute, Shanghai Jiao Tong University, Shanghai 200240, China}

\author{Peng Sun} 
\email{Corresponding author: 06260@njnu.edu.cn}
\affiliation{Department of Physics and Institute of Theoretical Physics, Nanjing Normal University, Nanjing, Jiangsu, 210023, China}

\author{Wei Wang}
\email{Corresponding author: wei.wang@sjtu.edu.cn}
\affiliation{INPAC, Shanghai Key Laboratory for Particle Physics and Cosmology, Key Laboratory for Particle Astrophysics and Cosmology (MOE), School of Physics and Astronomy, Shanghai Jiao Tong University, Shanghai 200240, China}

\author{Ji Xu}
\affiliation{INPAC, Shanghai Key Laboratory for Particle Physics and Cosmology, Key Laboratory for Particle Astrophysics and Cosmology (MOE), School of Physics and Astronomy, Shanghai Jiao Tong University, Shanghai 200240, China}
\affiliation{School of Physics and Microelectronics, Zhengzhou University, Zhengzhou, Henan 450001, China}

\author{Yi-Bo Yang}
\affiliation{CAS Key Laboratory of Theoretical Physics, Institute of Theoretical Physics, Chinese Academy of Sciences, Beijing 100190, China}
\affiliation{School of Fundamental Physics and Mathematical Sciences, Hangzhou Institute for Advanced Study, UCAS, Hangzhou 310024, China}
\affiliation{International Centre for Theoretical Physics Asia-Pacific, Beijing/Hangzhou, China}

\author{Jian-Hui Zhang}
\affiliation{Center of Advanced Quantum Studies, Department of Physics, Beijing Normal University, Beijing 100875, China}

\author{Qi-An Zhang}
\affiliation{Shanghai Key Laboratory for Particle Physics and Cosmology, Key Laboratory for Particle  Astrophysics and Cosmology (MOE), Tsung-Dao Lee Institute, Shanghai Jiao Tong University, Shanghai 200240, China}
 
\maketitle
  
%%%%%%%%%%%%%%%%%%%%%%%%%%%%%%%%%%%%%%%%%%%%%%%%%%%
%%%%%%%%%%%%%%%%%%%%%%%%%%%%%%%%%%%%%%%%%%%%%%%%%%%
%%%%%%%%%%%%%%%%%%%%%%%%%%%%%%%%%%%%%%%%%%%%%%%%%%%

\begin{widetext}

 %%%%%%%%%%%%%%%%%%%%%%%%%%%%%%%%%%%%%%%%%%%%%%%%%%%%%%%%%%%%%%
%\subsection{Renormalization Schemes}

\subsection{Matching kernel in Hybrid scheme}

 Here we present the one-loop matching in hybrid scheme for meson DA, where the matrix element used to renormalize the bare correlation is modified from the RIMOM matrix element. It reads
 \begin{align}\label{eq:hy_match1}
 C^{(1)}_{CT,Hybrid} &= \int dy' \left[\frac{P_z sin(u z_S)}{\pi u} - \frac{P_z sin(u_0 z_S)}{\pi u_0}  \right] \widetilde{q}^{(1)}(y')
     + \int dy' \left[e^{i(1-y')z_Sp^z_R} - 1 \right] \widetilde{q}^{(1)}(y')\delta(x-y)   \\ \nonumber
  	& -\int dy' \frac{P_z sin(u_0 z_S)}{\pi u_0} \left[e^{i(1-y')z_Sp^z_R} - 1 \right] \widetilde{q}^{(1)}(y'),
\end{align}
with 
 \begin{align}
 u=(1-y')p^z_R + (x-y)P_z,   \qquad u_0=(x-y)P_z.
 \end{align}
Alternatively, it can be written as
 \begin{align}
 C^{(1)}_{CT,Hybrid} =  C^{(1)}_{CT,RIMOM} + \int dy' \int \frac{dzP_z}{2\pi} \left[e^{i(1-y')z_Sp^z_R} - e^{i(1-y')z p^z_R} \right]   \widetilde{q}^{(1)}(y') \theta(|z| > z_S),
 \end{align}
where the $C^{(1)}_{CT,RIMOM}$ is the one-loop matching coefficient from RI/MOM scheme\cite{Liu:2018tox},

 \begin{align}
 C^{(1)}_{CT,RIMOM}\left(x,y,r, \frac{P_z}{\mu}, \frac{P_z}{p^z_R}\right) = \delta(x-y) + C^{(1)}_B\left(\Gamma,x,y,\frac{P_z}{\mu}\right) - C^{(1)}_{CT}\left(\Gamma,x,y,r,\frac{P_z}{p^z_R}\right)
  \end{align}
 
The bare matching coefficients are:
\begin{align}\label{eq:bare_matching}
C^{(1)}_B\left(\Gamma,x,y,\frac{P_z}{\mu}\right)=\frac{\alpha_s C_F}{2\pi}\left\{
\begin{array}{lc}
\left[H_1(\Gamma,x,y)\right]_{+(y)}				& x<0<y\\
\left[H_2(\Gamma,x,y,P_z/\mu)\right]_{+(y)}		& 0<x<y\\
\left[H_2(\Gamma,1-x,1-y,P_z/\mu)\right]_{+(y)}	& y<x<1\\
\left[H_1(\Gamma,1-x,1-y)\right]_{+(y)}			& y<1<x
\end{array}\right.
\end{align}
where
\begin{align}
H_1(\Gamma,x,y)&=\left\{
\begin{array}{ll}
\frac{1+x-y}{y-x}\frac{1-x}{1-y}\ln\frac{y-x}{1-x}+\frac{1+y-x}{y-x}\frac{x}{y}\ln\frac{y-x}{-x} & \Gamma=\gamma^z\gamma_5{\;\rm and\;}\gamma^t\\
\frac{1}{y-x}\frac{1-x}{1-y}\ln\frac{y-x}{1-x}+\frac{1}{y-x}\frac{x}{y}\ln\frac{y-x}{-x} & \Gamma=\gamma^z\gamma_\perp
\end{array} \right.\, ,\\
H_2\left(\Gamma,x,y,\frac{P_z}{\mu}\right)&=\left\{
\begin{array}{ll}
\frac{1+y-x}{y-x}\frac{x}{y}\ln\frac{4x(y-x)(P_z)^2}{\mu^2}+\frac{1+x-y}{y-x}\left(\frac{1-x}{1-y}\ln\frac{y-x}{1-x}-\frac{x}{y}\right) & \Gamma=\gamma^z\gamma_5\\
\frac{1+y-x}{y-x}\frac{x}{y}\left(\ln\frac{4x(y-x)(P_z)^2}{\mu^2}-1\right)+\frac{1+x-y}{y-x}\frac{1-x}{1-y}\ln\frac{y-x}{1-x} & \Gamma=\gamma^t\\
\frac{1}{y-x}\frac{x}{y}\ln\frac{4x(y-x)(P_z)^2}{\mu^2}+\frac{1}{y-x}\left(\frac{1-x}{1-y}\ln\frac{y-x}{1-x}-\frac{x}{y}\right) & \Gamma=\gamma^z\gamma_\perp
\end{array} \right. \, .
\end{align}
The counterterm of quasi-DA is determined by RI/MOM scheme. We use the renormalization factor for the quasi-PDF to renormalize the quasi-DA, which leads to the RI/MOM counterterm:
\begin{align}\label{eq:counterterm}
C^{(1)}_{CT}\left(\Gamma,x,y,r,\frac{P_z}{p^z_R}\right)=\left|\frac{P_z}{p^z_R}\right|\widetilde{q}^{(1)}\left(\Gamma,\frac{P_z}{p^z_R}(x-y)+1,r\right)_{+(y)}\,.
\end{align} 
$\widetilde{q}(\Gamma,x,r)$ is the real contribution of quasi-PDF at the RI/MOM subtraction scales $\mu_R$ and $p^z_R$, where $r = \mu^2_R/(p^z_R)^2$.
\begin{align}
\widetilde{q}^{(1)}(\gamma^z\gamma_5,x,r)=\frac{\alpha_s C_F}{2\pi}\left\{
\begin{array}{lc}
\frac{3r-(1-2x)^2}{2(r-1)(1-x)}-\frac{4x^2(2-3r+2x+4rx-12x^2+8x^3)}{(r-1)(r-4x+4x^2)^2}+\frac{2-3r+2x^2}{(r-1)^{3/2}(x-1)}\tan^{-1}\frac{\sqrt{r-1}}{2x-1} & x>1\\
\frac{1-3r+4x^2}{2(r-1)(1-x)}+\frac{-2+3r-2x^2}{(r-1)^{3/2}(1-x)}\tan^{-1}\sqrt{r-1} & 0<x<1\\
-\frac{3r-(1-2x)^2}{2(r-1)(1-x)}+\frac{4x^2(2-3r+2x+4rx-12x^2+8x^3)}{(r-1)(r-4x+4x^2)^2}-\frac{2-3r+2x^2}{(r-1)^{3/2}(x-1)}\tan^{-1}\frac{\sqrt{r-1}}{2x-1} & x<0
\end{array} \right. ,
\end{align}
\begin{align}
\widetilde{q}^{(1)}(\gamma^t,x,r)=\frac{\alpha_s C_F}{2\pi}\left\{
\begin{array}{lc}
\frac{-3r^2+13rx-8x^2-10rx^2+8x^3}{2(r-1)(x-1)(r-4x+4x^2)}+\frac{-3r+8x-rx-4x^2}{2(r-1)^{3/2}(x-1)}\tan^{-1}\frac{\sqrt{r-1}}{2x-1} & x>1\\
\frac{-3r+7x-4x^2}{2(r-1)(1-x)}+\frac{3r-8x+rx+4x^2}{2(r-1)^{3/2}(1-x)}\tan^{-1}\sqrt{r-1} & 0<x<1\\
-\frac{-3r^2+13rx-8x^2-10rx^2+8x^3}{2(r-1)(x-1)(r-4x+4x^2)}-\frac{-3r+8x-rx-4x^2}{2(r-1)^{3/2}(x-1)}\tan^{-1}\frac{\sqrt{r-1}}{2x-1} & x<0
\end{array} \right. ,
\end{align}
\begin{align}
\widetilde{q}^{(1)}(\gamma^z\gamma_\perp,x,r)=\frac{\alpha_s C_F}{2\pi}\left\{
\begin{array}{lc}
\frac{3}{2(1-x)}+\frac{r-2x}{(r-1)(r-4x+4x^2)}+\frac{r-2x+rx}{(r-1)^{3/2}(1-x)}\tan^{-1}\frac{\sqrt{r-1}}{2x-1} & x>1\\
\frac{1-3r+2x}{2(r-1)(1-x)}+\frac{r-2x+rx}{(r-1)^{3/2}(1-x)}\tan^{-1}\sqrt{r-1} & 0<x<1\\
-\frac{3}{2(1-x)}-\frac{r-2x}{(r-1)(r-4x+4x^2)}-\frac{r-2x+rx}{(r-1)^{3/2}(1-x)}\tan^{-1}\frac{\sqrt{r-1}}{2x-1} & x<0
\end{array} \right. .
\end{align}

 \end{widetext}

 {\subsection{Renormalization and matching using the hybrid scheme}}
In this work, we have performed the renormalization and matching in the hybrid scheme~\cite{Ji:2020brr}. This scheme takes advantages of both the RI/MOM or ratio type renormalization and the gauge-link mass subtraction scheme in that the former cancels discretization effects at short distances whereas the latter avoids introducing extra nonperturbative effects at large distances at the renormalization stage. 

\begin{figure}[!th]
\begin{center}
\includegraphics[width=0.47\textwidth]{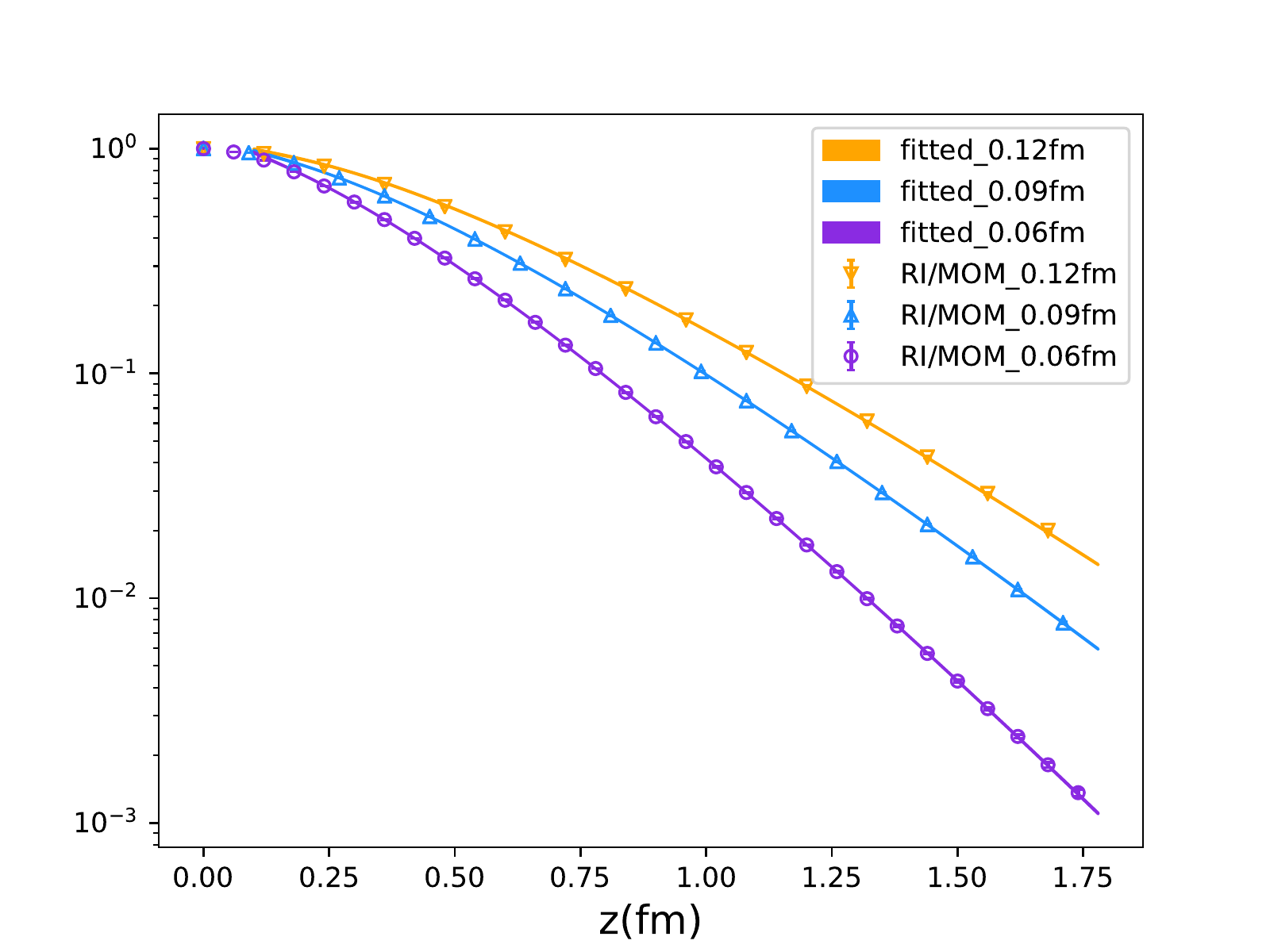}
\caption{Fit of RI/MOM renormalization factor with different lattice spacings. }
\label{fig:fitnpr}
\end{center}
\end{figure}

\begin{figure}[!th]
\begin{center}
\includegraphics[width=0.47\textwidth]{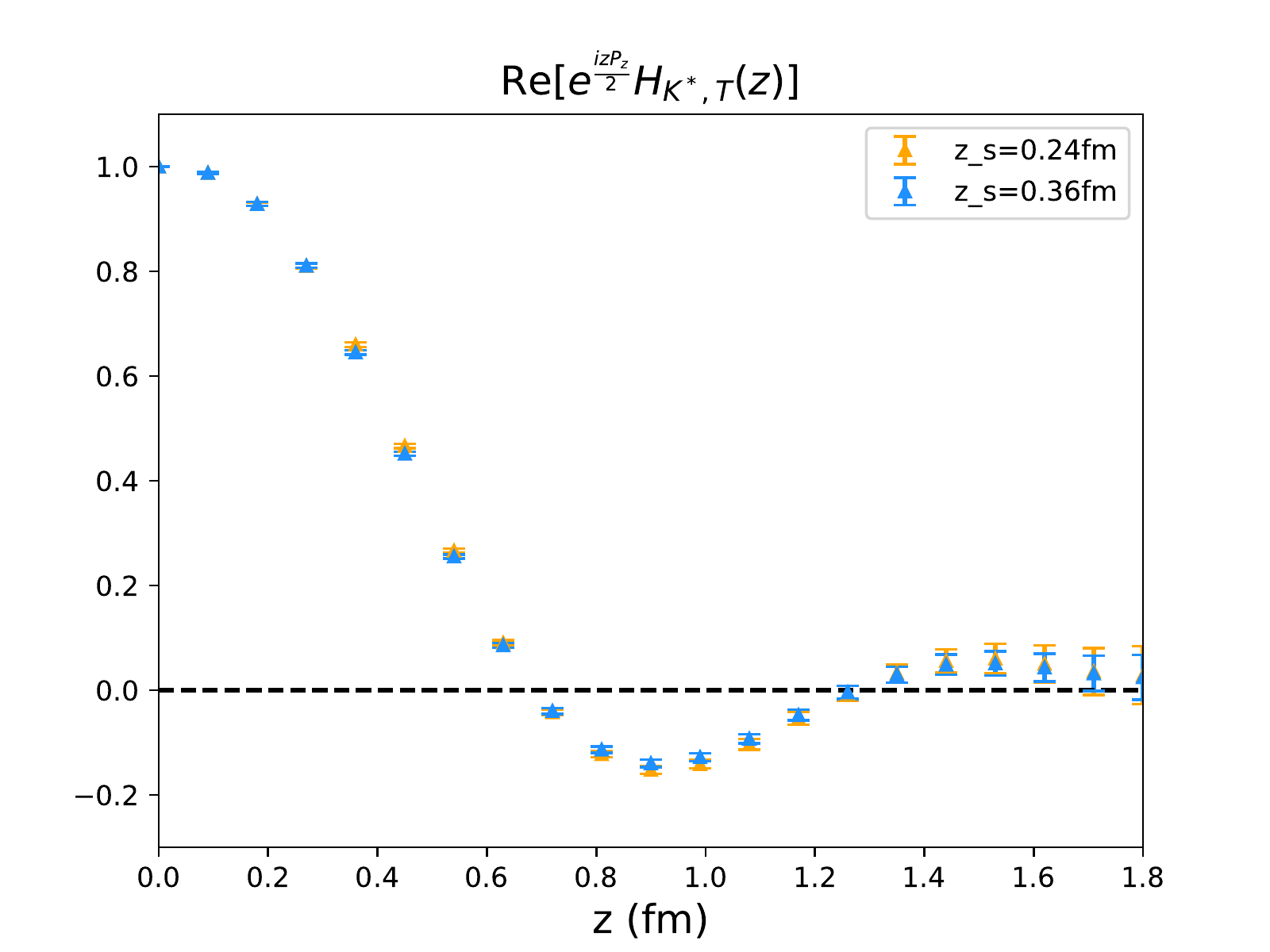}
\caption{The renormalized quasi-DA of transversely-polarized $K^*(a=0.09fm,P_z=2.15GeV)$ in coordinate space with two different values of $z_s$ in Hybrid approach.}
\label{fig:zs_dp}
\end{center}
\end{figure}

The renormalization in the hybrid scheme has been given in Eq.~(4) of the main text, where the mass counterterm can be written as 
\begin{equation}
\delta m=m_{-1}/a+m_0,
\end{equation}
with $m_0\sim \Lambda_{\rm QCD}$, and can be extracted in practice from the asymptotic behavior of various hadronic matrix elements. Here we choose to use the RI/MOM renormalization factor at large distance and perform a fit to the quasi-LF correlation from moderate to large $z$ using the following form:
\begin{eqnarray}\label{eq:firnpr}
	C_2 = C_0e^{-(1+c_2 {\rm ln }\frac{z}{z_s})\left[\left(\frac{m_{-1}}{a} + \frac{m_0}{z_0}\right)\frac{z}{z_0} + c_1{\rm ln}\frac{z}{z_0}  \right]},
\end{eqnarray}
where an extra logarithmic term in front of the square bracket is introduced  to  account for the dependence of strong coupling on the distance, $z_0=1~$fm for dimensionless ratios, and $z_s = 0.24~$fm is the starting point of the fit. The fit plot is shown in Fig.~\ref{fig:fitnpr}.
 To account for the systematic effects from $z_s$, we have chosen two different values  
(0.24fm and 0.36fm) for $z_s$.  As shown in the Fig.~\ref{fig:zs_dp}, they tend to have small effects on the renormalized quasi-DA.
To ensure the rigor of the results, the differences are treated as an estimate of the $z_s$ dependence.

\begin{figure}[tbp]
\begin{center}
\includegraphics[width=0.47\textwidth]{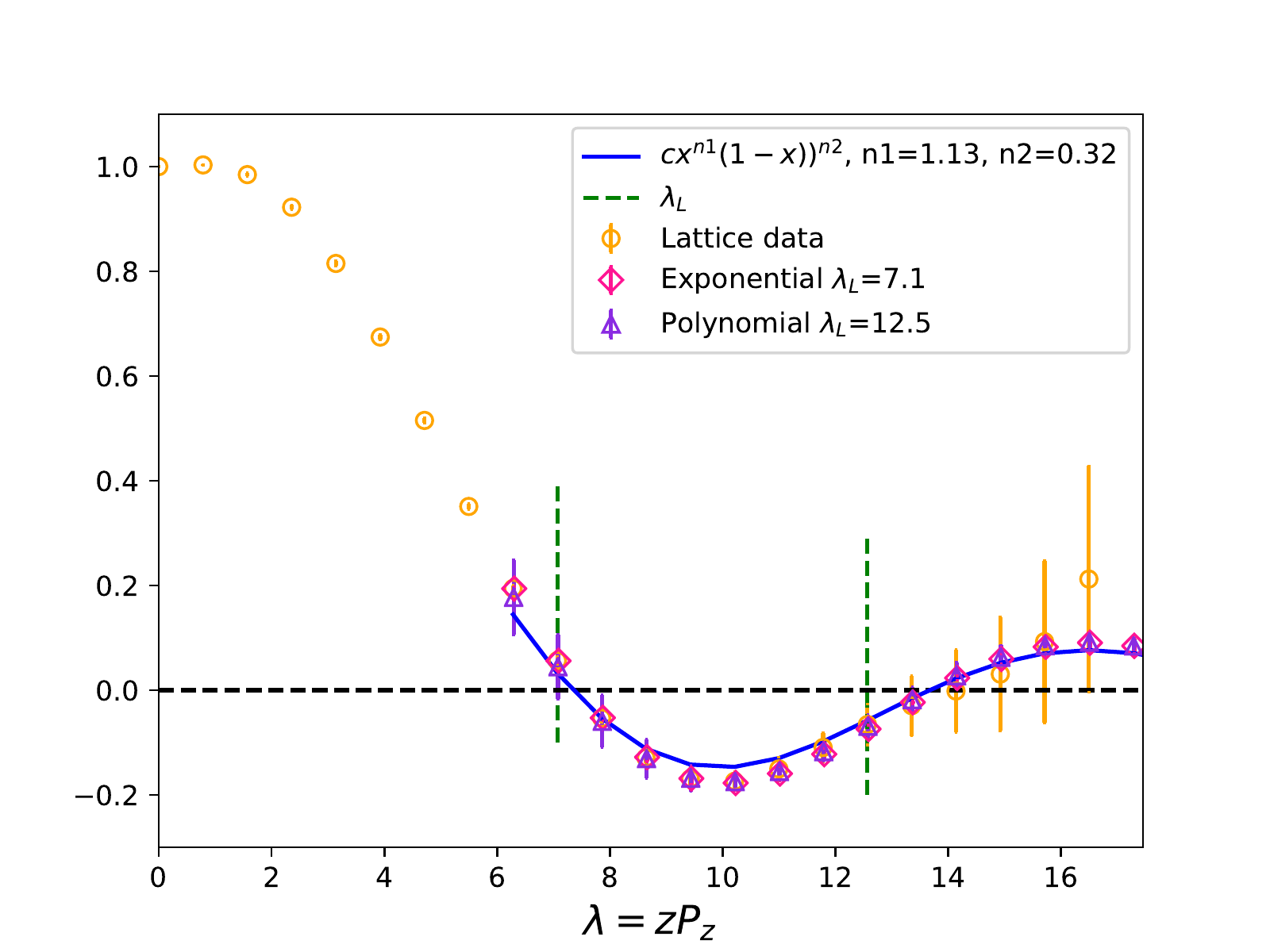}
\caption{Comparison of different extrapolations and lattice data.}\label{fig:extra}
\end{center}
\end{figure}

%\subsection{Extrapolation}
%With Hybrid approach, we will use a physics-based extrapolation form in large z area. 
In practice, %The extrapolation form are base on the asymptotic behavior of parton distributions at $x \to 0/1$: $x^a(1-x)^b$,
%after Fourier transformation and expansion at large $\lambda(zP_z)$, the polynomial extrapolation form  we used in large $\lambda$ are:
we take the following two extrapolation forms~\cite{Ji:2020brr}, one  exponential and the other algebraic
\begin{align}
         \tilde H(z, P_z) &= \left[\frac{c_1}{(-i\lambda)^a} + e^{i\lambda}\frac{c_2}{(i \lambda)^b}\right]e^{-\frac{\lambda}{\lambda_0}}, \nonumber\\
	\tilde H(z, P_z) &= \frac{c_1}{(-i\lambda)^a} + e^{i\lambda}\frac{c_2}{(i \lambda)^b},
\end{align}
and use their difference as an estimate of systematic errors from extrapolation. %The algebraic form above follows from the asymptotic behavior of momentum distribution at $x\to {0, 1}$~\cite{Ji:2020brr}. 
Fig.~\ref{fig:extra} shows the comparison of two extrapolation forms, which are consistent with each other within errors. We have also tested that the same behavior  is achieved when the fitting range is varied.

As for the hybrid matching, it turns out to approach the RI/MOM one when $p_z=0$. 
\\

 %%%%%%%%%%%%%%%%%%%%%%%%%%%%%%%%%%%%%%%%%%%%%%%%%%%%%%%%%%%%%%
\subsection{Comparison of constant fit with two-state fit}
As shown in Fig.1 of the main text, we perform a phase rotation $e^{izP_z/2}$ to the renormalized correlations, so that the imaginary part directly reflects the flavor SU(3) asymmetry. The reliability of the two-state fit in Eq. (10) of the main text largely depends on whether the energy of the excited state can be  well determined. However, after the phase rotation, weights of the excited state energy in the real and imaginary parts may be different, therefore the two-state fit including excited energy is unstable. 

\begin{figure}[!th]
\begin{center}
\includegraphics[width=0.4\textwidth]{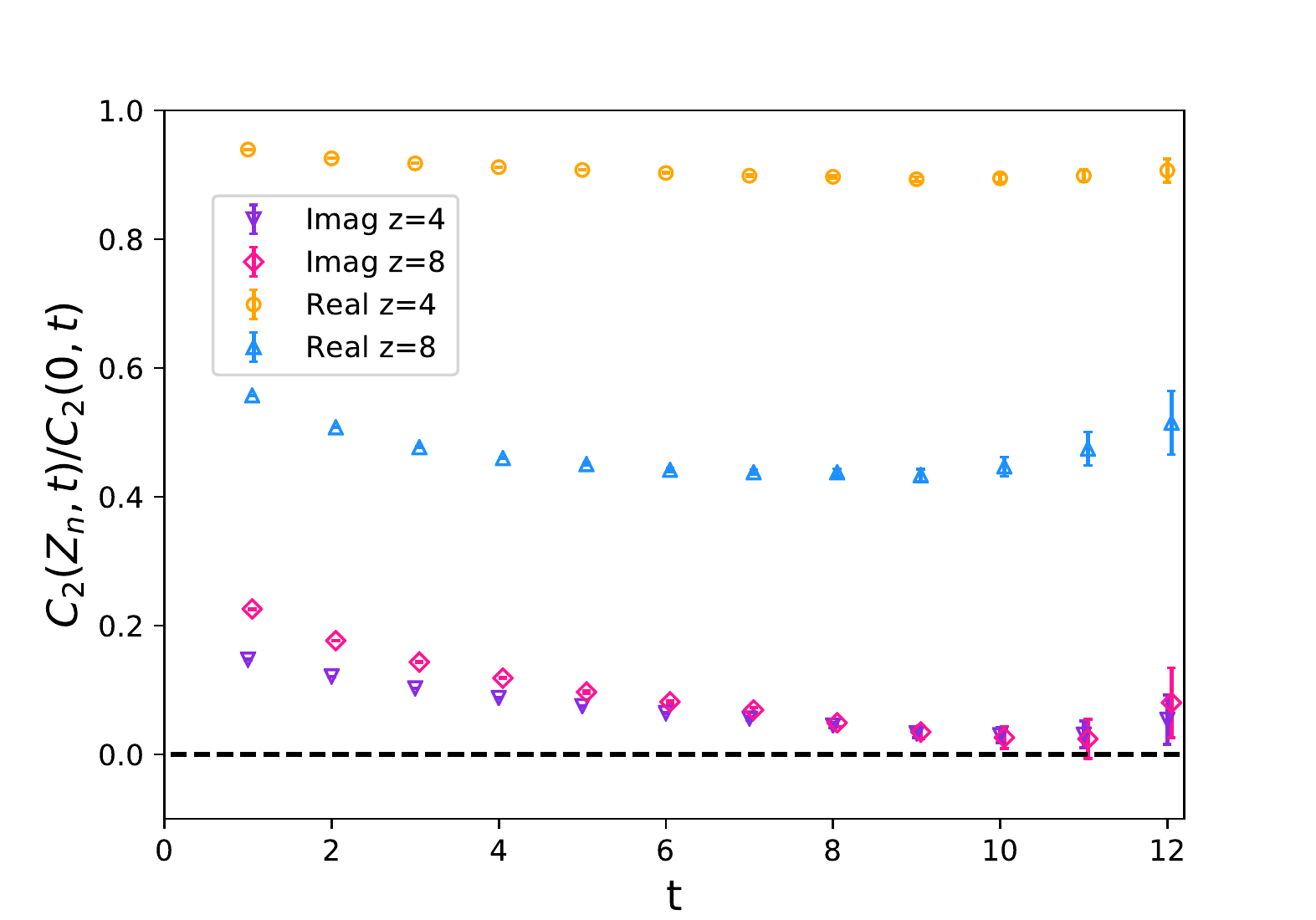}
\caption{Trends in the ratios $C^m_2(z,\vec P, t) / C^m_2(z=0,\vec P, t)$  of the real and imaginary parts over time at lattice spacing $a=0.06~$fm. Take $z=4$ and $z=8$ as examples.}\label{fig:fixt_support2}
\end{center}
\end{figure}

An alternative plan is to obtain the ground state matrix element $H_{V,m}^{b}(z)$ in Eq. (10) of the main text by the constant fit (set $\Delta E =0$) to eliminate the influence of excited state energy.  When $t$ is large enough, the excited states contamination parameterized by $c_{m}(z)$ and $\Delta E$ are indeed suppressed exponentially, and the ratio $C^m_2(z,\vec P, t) / C^m_2(z=0, \vec P, t)$ approaches the ground state matrix element $H_{V,m}^{b}(z)$. In Fig.  \ref{fig:fixt_support2} we show as an example the trend of the above ratio over time at lattice spacing $a=0.06~$fm. As can be seen from the figure, the imaginary part decays to the ground state more slowly than the real part. We make a more conservative choice for the region $t \geq 9a =0.54~$fm for the constant fit.

\begin{figure}[!th]
\begin{center}
\includegraphics[width=0.47\textwidth]{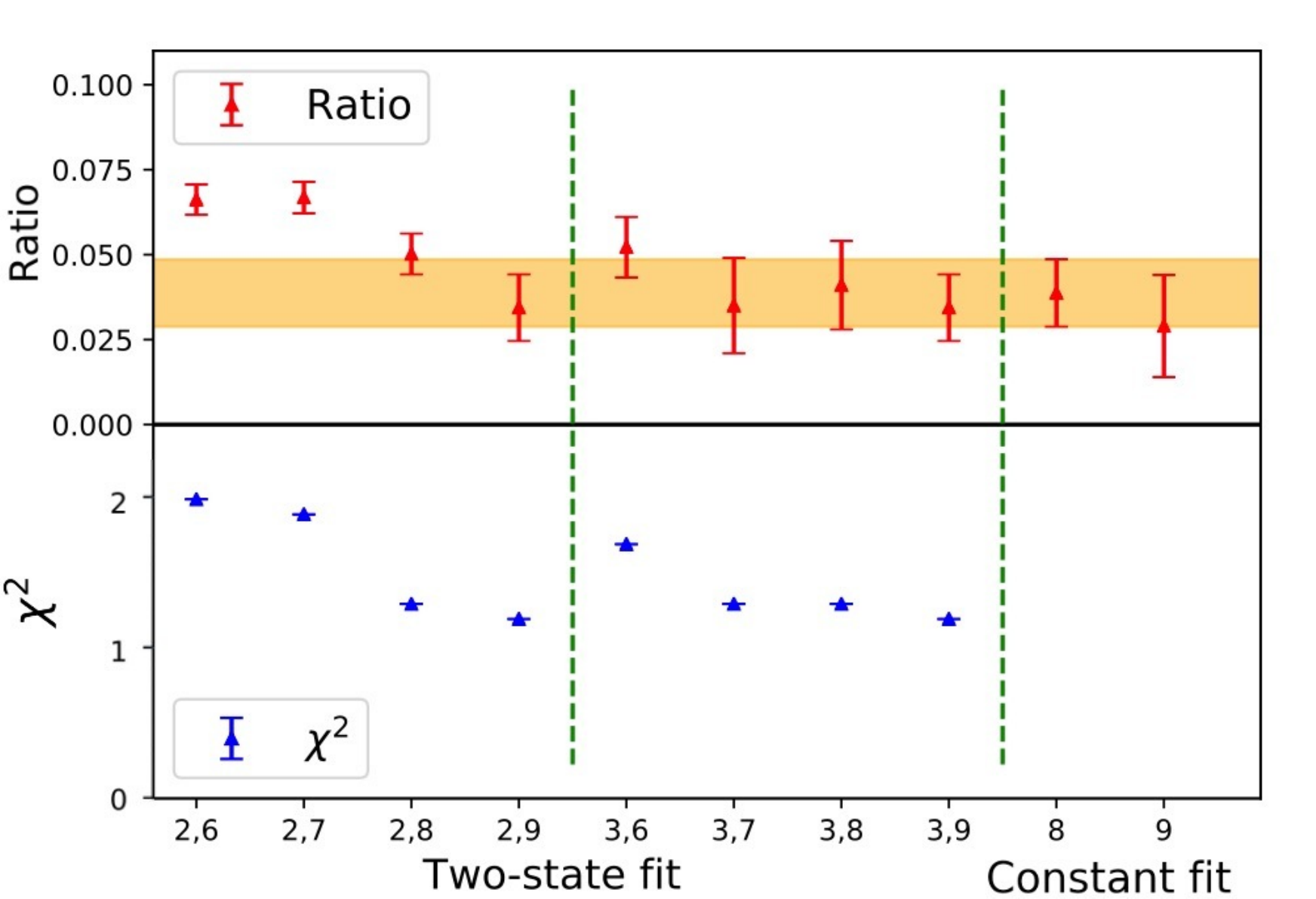}
\caption{The upper half plane shows the comparison of two-state fitted ratio  $H_{V,m}^{b}(z)$ and constant fitted ones.  The lower half of the plane shows the $\chi^2$ of the two-state fit.}\label{fig:fixt_support3}
\end{center}
\end{figure}

Fig. \ref{fig:fixt_support3} shows the comparison of  $H_{V,m}^{b}(z)$ derived by two-state fit and constant fit. The paired number $(2,6)$ in the two-state fit implies that the fit range for the local correlation functions is $t \geq 2$  and for the non-local correlation functions is $t \geq 6$. The two-state fitted $H_{V,m}^{b}(z)$ will become stable only when the starting point of the non-local correlation function fit range becomes very large. In this case,  the two-state fitted ratios $H_{V,m}^{b}(z)$ are consistent with the constant fitted ones within statistical uncertainties.

 \subsection{Discretization error}

\begin{figure}[!th]
\begin{center}
\includegraphics[width=0.47\textwidth]{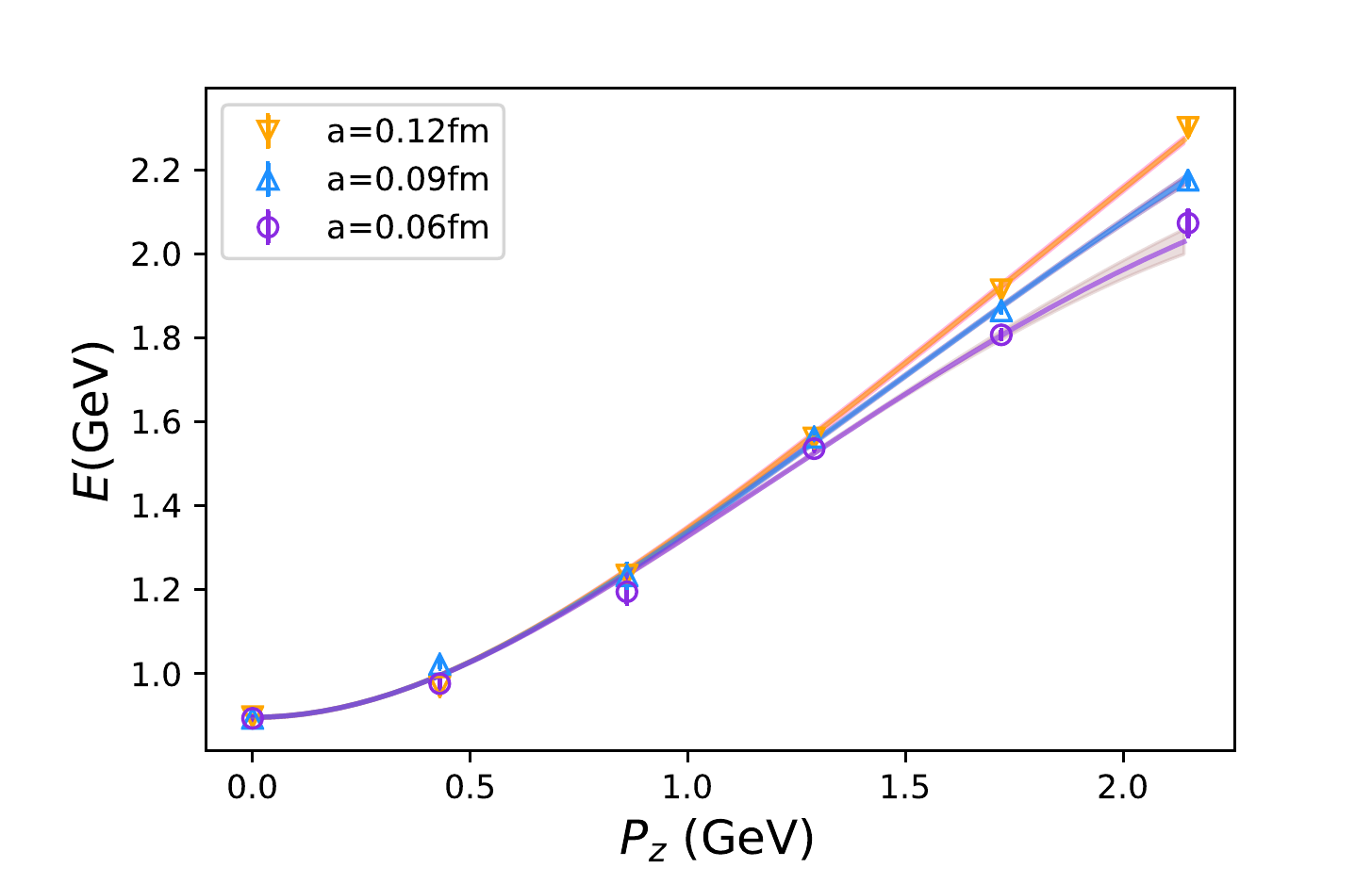}
\includegraphics[width=0.47\textwidth]{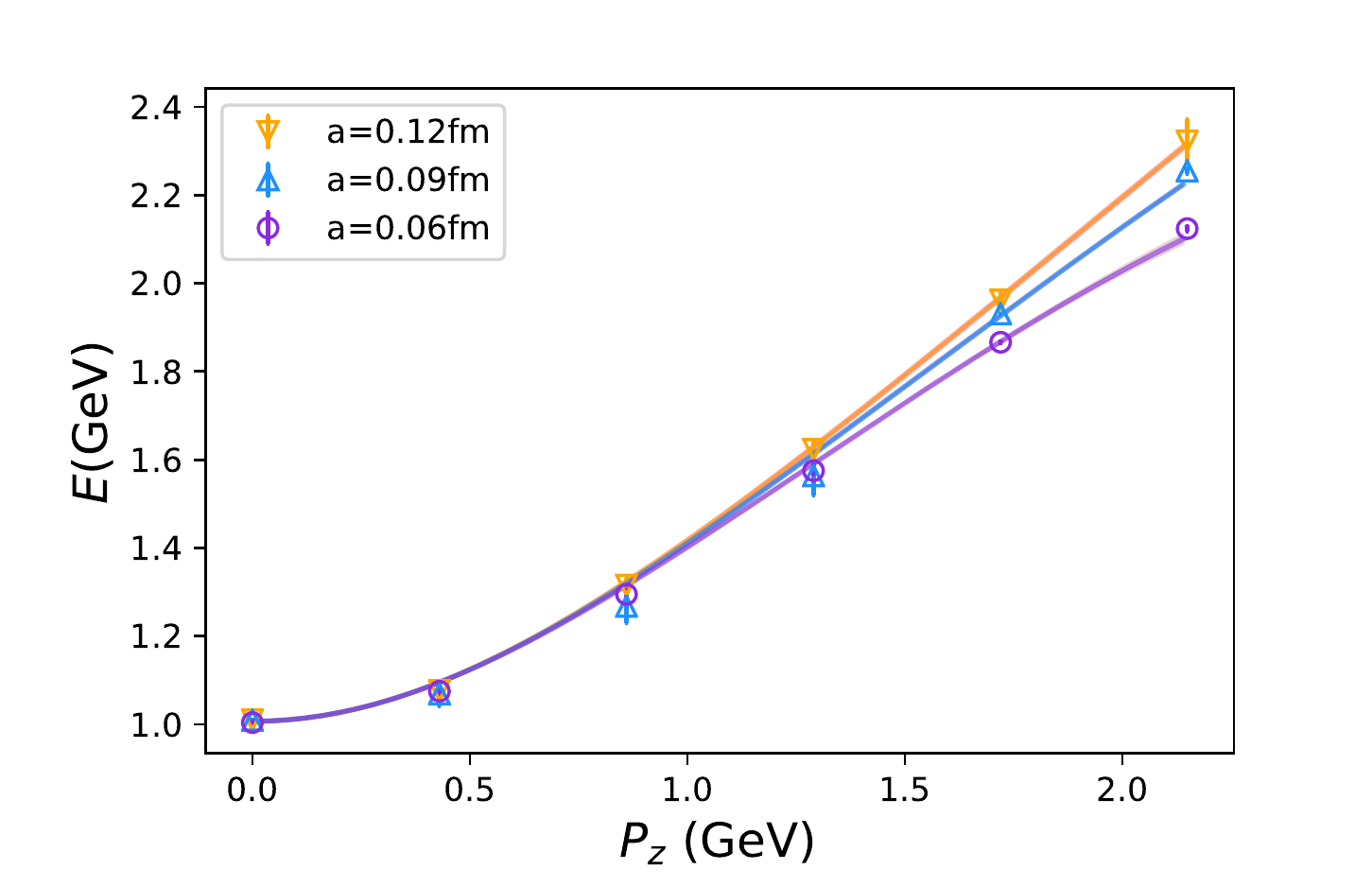}
\caption{Dispersion relation for the $K^*$ (upper panel) and $\phi$ (lower panel) meson.}\label{fig:dispersion_relation}
\end{center}
\end{figure}

Based on the two-point function at $z=0$, one can extract the effective mass through a two-state fit. 
We combine the longitudinally-polarized and transversely-polarized data to give the dispersion relation for $K^*$ and $\phi$ in Fig.~\ref{fig:dispersion_relation}. The parametrized form is given as:
\begin{eqnarray}
E^2 = m^2 + c_2(P_z)^2+c_3(P_z)^4a^2.
\end{eqnarray}
The  $c_{3, K^*}=-0.178\pm0.024$ and  $c_{3, \phi}=-0.158\pm0.010$  reflects the discretization effects, and within $2 \sigma$ deviation,  $c_{2, K^*}=1.026\pm0.016$, $c_{2, \phi}=1.009\pm0.009$ are consistent with the speed of light.  {It shows that the dispersion relation is satisfied up to ${\cal O}(p^4a^2)$ correction.}

%%%%%%%%%

 {We also made a test on the two continuum extrapolation procedures (in coordinate or momentum space) for  transversely-polarized  $K^*$ DA  with $P_z=2.15$ GeV.  As shown in Fig.~\ref{fig:qda_limit} in this reply, the two extrapolations actually give consistent results for the extracted quasi DAs, within the statistical uncertainty.}

\begin{figure}[!th]
\begin{center}
\includegraphics[width=0.47\textwidth]{qda.pdf}
\includegraphics[width=0.47\textwidth]{qda_limit.pdf}
\caption{The two choices of continuum extrapolation. The left diagram is the  continuum extrapolation in coordinate space, and 
the right one is the comparison of two extrapolation types shown in momentum space. The red data in the right diagram is continuum extrapolated 
in coordinate space then Fourier transform to momentum space. The blue data in the right diagram is continuum extrapolated 
in momentum space.}\label{fig:qda_limit}
\end{center}
\end{figure}

%%%%%%%%%

 %%%%%%%%%%%%%%%%%%%%%%%%%%%%%%%%%%%%%%%%%%%%%%%%%%%%%%%%%%%%%%
\subsection{Quasi-DA in coordinate space  {and Fourier transform}}
We perform a phase rotation $e^{izP_z/2}$ for the two-point correlation function in Fig.~1 of the main text. 
For comparison, we show the two-point correlation function of transversely-polarized $K^*$ without phase rotation in Fig.~\ref{fig:qDAkstarT}.
The upper panel is the real part and the lower panel is the imaginary part.
\begin{figure}[!th]
\begin{center}
\includegraphics[width=0.47\textwidth]{ks_T_Real.pdf}
\includegraphics[width=0.47\textwidth]{ks_T_Imag.pdf}
\caption{The quasi-DA for the transversely-polarized $K^*$ in coordinate space with $P_z=2.15$ GeV. }\label{fig:qDAkstarT}
\end{center}
\end{figure}

One has to Fourier transforming the renormalized correlations to momentum space.  For finite hadron momentum, the available lattice data of quasi-LF correlations always end up at finite $\lambda_L = z_L P_z$ while one needs the correlations at all quasi-light-front distances $\lambda$ to reconstruct the momentum distribution. Here we follow the extrapolation strategy proposed in Ref.~\cite{Ji:2020brr} based on generic properties of coordinate space correlations. At finite momentum, the quasi-LF correlation in general has a finite correlation length and exhibits an exponential decay (usually associated with a power decay as well). When the momentum gets larger, the correlation length (in $\lambda$ space) also gets larger and the exponential decay behavior is much extended in the quasi-light-front correlation. Thus it is anticipated  that the decay behavior follows more like an algebraic law rather than an exponential one as $\lambda\sim \lambda_L$, where the power law behavior is also consistent with the asymptotic Regge behavior of momentum distribution~\cite{Regge:1959mz}.

  %%%%%%%%%%%%%%%%%%%%%%%%%%%%%%%%%%%%%%%%%%%%%%%%%%%%%%%%%%%%%%
 \subsection{Numerical results of LCDAs}
 The specific numerical results of LCDAs of $K^*$ and $\phi$ are given in Table.~\ref{tab:num_DA}. The data in the region of $x<0.1$ and $x>0.9$ are not realiable, 
 so we do not give this part of the data. In phenomenological calculations, the LCDAs are widely expanded by Gegenbauer polynomials as:
\begin{align}
	\phi(x) = 6x(1-x)\left[1 + \sum^\infty_{n}  a_nC_n^{3/2}(2x-1) \right] , 
\end{align}
where $a_n$ is Gegenbauer moments and $C_n$ is Gegenbauer polynomials. One can extract 
Gegenbauer moments from LCDAs following the form\cite{Braun:2016wnx}:
\begin{align}
	a_n = \frac{2(2n+3)}{3(n+1)(n+2)}\int^1_0 dx C_n^{3/2}(2x-1) \phi(x).
\end{align}
The Gegenbauer moments extracted from our results of LCDAs are presented in Table.~\ref{tab:moment}. \Red{Due to the requirement of the charge-conjugation invariance, the odd Gegenbauer moments such as $a_1$ of the $\phi$ meson are identically zero.  }

\begin{table*}[]
\caption{The numerical results of LCDAs of $K^*$ and $\phi$,  where $x$ is the momentum fraction and the error in
the bracket including the statistical error and the systematic error from the renormalization scale, $z_S$ and extrapolation form dependences in the Hybrid approach. }
\label{tab:num_DA}
\centering
\begin{tabular}{cccccccccc}
\hline
\hline
~~~~~$x$ ~~~~&~~~~~ $\phi_{K^*,L}$ ~~~~&~~~~~  $\phi_{K^*,T}$   ~~~~&~~~~~   $\phi_{\phi^*,L}$  ~~~~&~~~~~  $\phi_{\phi^*,T}$  ~~~~~ \\
\hline
%~~0.05~~&~~     0.30(11)~~&~~   0.693(44)~~&~~  0.33(11)~~&~~   0.535(60)~~\\
0.1\,\,\,~~&~~      0.532(80)~~&~~  0.928(59)~~&~~  0.577(71)~~&~~  0.732(52)~~\\
0.15~~&~~          0.738(46)~~&~~  1.067(68)~~&~~  0.773(47)~~&~~  0.888(50)~~\\
0.2\,\,\,~~&~~      0.920(19)~~&~~  1.150(65)~~&~~  0.946(22)~~&~~  1.001(40)~~\\
0.25~~&~~          1.080(22)~~&~~  1.202(59)~~&~~  1.099(19)~~&~~  1.087(33)~~\\
0.3\,\,\,~~&~~      1.214(41)~~&~~  1.230(48)~~&~~  1.229(42)~~&~~  1.155(35)~~\\
0.35~~&~~          1.320(57)~~&~~  1.241(38)~~&~~  1.330(60)~~&~~  1.214(38)~~\\
0.4\,\,\,~~&~~      1.397(68)~~&~~  1.237(37)~~&~~  1.399(70)~~&~~  1.243(47)~~\\
0.45~~&~~          1.444(76)~~&~~  1.222(40)~~&~~  1.440(76)~~&~~  1.256(59)~~\\
0.5\,\,\,~~&~~      1.462(78)~~&~~  1.199(46)~~&~~  1.454(77)~~&~~  1.261(63)~~\\
0.55~~&~~          1.448(76)~~&~~  1.165(53)~~&~~  1.439(76)~~&~~  1.256(59)~~\\
0.6\,\,\,~~&~~      1.401(68)~~&~~  1.127(57)~~&~~  1.397(70)~~&~~  1.243(47)~~\\
0.65~~&~~          1.325(57)~~&~~  1.083(57)~~&~~  1.328(60)~~&~~  1.213(38)~~\\
0.7\,\,\,~~&~~      1.223(40)~~&~~  1.034(54)~~&~~  1.227(42)~~&~~  1.153(35)~~\\
0.75~~&~~          1.096(24)~~&~~  0.978(50)~~&~~  1.096(19)~~&~~  1.085(33)~~\\
0.8\,\,\,~~&~~      0.945(38)~~&~~  0.910(54)~~&~~  0.943(22)~~&~~  0.998(41)~~\\
0.85~~&~~          0.770(68)~~&~~  0.827(66)~~&~~  0.769(47)~~&~~  0.885(50)~~\\
0.9\,\,\,~~&~~      0.57(10)\,\,\,~~&~~   0.715(82)~~&~~  0.572(70)~~&~~  0.728(53)~~\\
%~~0.95~~&~~     0.34(12)~~&~~   0.56(10)~~&~~   0.32(11)~~&~~   0.531(62)~~\\
\hline
\end{tabular}
\end{table*}

\begin{table*}[]
\caption{The Gegenbauer moments extracted from LCDAs, the numbers in the two brackets following the central value are statistical error,  and also the systematic error from the renormalization scale, $z_S$ and extrapolation form dependences in the Hybrid approach.}
\label{tab:moment}
\centering
\begin{tabular}{cccccccccc}
\hline
\hline
Gegenbauer moments &  $a_1$      &  $a_2$    &   $a_4$   \\
\hline
$K^*,  L$         & {\color{red} -0.005(07)(07)}  ~~ &  0.015(10)(08)   ~~~&   0.013(09)(09) \\
$K^*,  T$         &  -0.074(06)(07)  ~~ &  0.181(07)(12)   ~~~&   0.064(07)(06)  \\ 
$\phi, L$          &   ~~~$--$ ~~  & ~0.018(09)(09)  ~~~ &  0.007(10)(20)        \\
$\phi, T$          &   ~~~$--$ ~~  & ~0.128(03)(21)  ~~~ &  0.044(04)(08)         \\
\hline
\end{tabular}
\end{table*}

 %%%%%%%%%%%%%%%%%%%%%%%%%%%%%%%%%%%%%%%%%%%%%%%%%%%%%%%%%%%%%%